\documentclass[lettersize,journal]{IEEEtran}
\IEEEoverridecommandlockouts
\usepackage{cite}
\usepackage{bm}
\usepackage{amsmath,amssymb,amsfonts}
\usepackage{amsthm}
\usepackage{algorithm}
\usepackage{algpseudocode}
\usepackage{graphicx}
\usepackage{amsfonts}
\usepackage{amssymb}
\usepackage{textcomp}
\usepackage[T1]{fontenc}
\usepackage{enumitem}
\usepackage{longtable} 
\usepackage{supertabular} 
\usepackage{xcolor}
\usepackage{subcaption}
\usepackage{tabularx}
\usepackage{comment}
\usepackage{booktabs}
\usepackage{mathrsfs} 
\usepackage{float} 
\usepackage{booktabs} 
\usepackage{mathtools}
\mathtoolsset{centercolon}
\usepackage{multirow}

\newtheorem{definition}{Definition}
\newtheorem{theorem}{Theorem}

\newtheorem{assumption}{Assumption}
\newtheorem{remark}{Remark}

\allowdisplaybreaks
\def\BibTeX{{\rm B\kern-.05em{\sc i\kern-.025em b}\kern-.08em
    T\kern-.1667em\lower.7ex\hbox{E}\kern-.125emX}}

\allowdisplaybreaks
\def\BibTeX{{\rm B\kern-.05em{\sc i\kern-.025em b}\kern-.08em
    T\kern-.1667em\lower.7ex\hbox{E}\kern-.125emX}}
\begin{document}

\title{
Neural Cooperative Reach-While-Avoid Certificates for Interconnected Systems

\author{Jingyuan Zhou, Haoze Wu, and Kaidi Yang}%
\thanks{Jingyuan Zhou and Kaidi Yang are with the Department of Civil and Environmental Engineering, National University of Singapore, Singapore 119077. Email:{jingyuanzhou@u.nus.edu,
kaidi.yang@nus.edu.sg}. (\emph{Corresponding Author: Kaidi Yang})}
\thanks{Haoze Wu is with the Department of Computer Science, Amherst College, P.O. Box 5000, Amherst, MA 01002.}
}

\maketitle

\begin{abstract}
Providing formal guarantees for neural network-based controllers in large-scale interconnected systems remains a fundamental challenge. In particular, using neural certificates to capture cooperative interactions and verifying these certificates at scale is crucial for the safe deployment of such controllers. However, existing approaches fall short on both fronts. To address these limitations, we propose neural cooperative reach-while-avoid certificates with Dynamic-Localized Vector Control Lyapunov and Barrier Functions, which capture cooperative dynamics through state-dependent neighborhood structures and provide decentralized certificates for global exponential stability and safety. Based on the certificates, we further develop a scalable training and verification framework that jointly synthesizes controllers and neural certificates via a constrained optimization objective, and leverages a sufficient condition to ensure formal guarantees considering modeling error. To improve scalability, we introduce a structural reuse mechanism to transfer controllers and certificates between substructure-isomorphic systems. The proposed methodology is validated with extensive experiments on multi-robot coordination and vehicle platoons. Results demonstrate that our framework ensures certified cooperative reach-while-avoid while maintaining strong control performance.
\end{abstract}


\section{Introduction}
Large-scale interconnected systems are prevalent in real-world applications such as intelligent transportation~\cite{ZHOU2024104885}, robotics~\cite{ren2023vector}, and smart grids~\cite{silva2021string}. Advances in sensing and communication technologies have made these systems increasingly complex and tightly coupled. A central challenge in controlling such systems is ensuring both stability and safety. 

Control strategies for ensuring stability and safety in interconnected systems can be broadly categorized into model-based and learning-based approaches. Model-based controllers, such as barrier Lyapunov function-based control~\cite{zhang2022adaptive}, model predictive control~\cite{liu2017distributed}, and sliding mode control~\cite{guo2016distributed}, explicitly incorporate safety and stability into control objectives and constraints. However, these approaches often rely on accurate system models and may become ineffective when facing the complex and uncertain dynamics typical of large-scale interconnected systems. To overcome these limitations, learning-based controllers~\cite{fuentes2020adaptive,gu2021safety,zhang2022barrier,wei2022stability,cai2023fixed,liu2023reinforcement,zhou2024enhancing}, particularly those based on neural networks, have gained increasing popularity due to their ability to manage complex dynamics and uncertainties. However, the black-box nature of neural networks makes it challenging to provide guarantees for stability and safety. In most existing neural network-based controllers, safety and stability are treated as soft constraints (e.g., included in the loss or reward function), without offering rigorous theoretical assurances.

To the best of our knowledge, few studies have addressed the challenge of verifying and ensuring stability and safety in neural network-based controllers for large-scale interconnected systems. One notable attempt is~\cite{zhang2025gcbf+}, which incorporates neural graph control barrier function certificates with nominal controllers for large-scale multi-agent systems. However, this approach does not account for cooperation among agents and lacks formal performance verification. Another relevant work~\cite{zhou2025synthesis} considers neural string stability certificates and employs neural network verification tools to guarantee system-level string stability. However, it focuses solely on stability while neglecting safety, and does not consider the limited communication range inherent in interconnected systems. Several studies have attempted to provide formal guarantees for local stability and safety~\cite{dai2021lyapunov,yanglyapunov,zhang2023compositional,mandal2024safe,mandal2024formally}. For example,~\cite{mandal2024formally} leverages Lyapunov barrier certificates to formally verify deep reinforcement learning (RL) controllers, demonstrating safe and reliable training in aerospace applications. However, these approaches are not readily extensible for large-scale interconnected systems for two reasons. First, these approaches often overlook inter-agent cooperation and interaction, and thus the resulting controllers may be overly conservative. Second, it is computationally inefficient to train and verify large-scale interconnected systems. Consequently, formally guaranteeing stability and safety in large-scale interconnected systems remains an open and challenging problem.

\emph{Statement of Contribution}. In this paper, we focus on a special class of control problems in interconnected systems, namely, multi-agent reach-while-avoid, whereby a set of agents seek to reach a goal set (encoded via global asymptotic stability) while avoiding collisions. Our contributions are three-fold. First, we initiate neural \emph{cooperative} reach-while-avoid (Co-RWA) certificates with dynamic-localized vector control Lyapunov and barrier functions that provide theoretical guarantees of exponential global stability and safety for interconnected systems with cooperation consideration and dynamic neighborhoods. 
Second, we formulate a scalable verification framework by encoding the proposed neural dynamic-localized Lyapunov and barrier functions into a neural network verification problem. Specifically, we propose a sufficient condition for verification considering dynamics modeling errors and design a counterexample-guided synthesis (CEGIS) framework to verify and train the neural controllers. In the CEGIS framework, an initial controller is trained using RL or by imitating traditional model-based controllers, and is subsequently fine-tuned using counterexamples generated by a verification process that identifies violations of the certification conditions. Compared with existing CEGIS-based frameworks that tend to focus on low-dimensional single-agent systems, we enhance the scalability to large-scale systems by proposing a mechanism to transfer synthesized controllers and neural certificates from small systems to large ones. 
Third, we validate the proposed framework on multi-robot coordination and vehicle platoon scenarios. Simulation results demonstrate that our approach ensures Co-RWA for large-scale interconnected systems while maintaining control performance.

The rest of the paper is organized as follows. Section~\ref{sec: Related Work} reviews related work. Section~\ref{sec:Neural Cooperative Reach-While-Avoid Certificates} introduces the proposed neural Co-RWA certificates. Section~\ref{sec:framework} presents the scalable synthesis and verification framework. Section~\ref{sec:Numerical Simulation} reports numerical simulations in different scenarios. Section~\ref{sec:conclusion} concludes the paper.

\section{Related Work}
\label{sec: Related Work}
This work makes contributions in two main research areas: (i) controller design for large-scale interconnected systems, and (ii) neural certificates and neural network verification.

\noindent\textbf{Control of Large-Scale Interconnected Systems.}
Model-based control strategies for interconnected systems have been extensively studied using methods such as distributed model predictive control~\cite{liu2017distributed} and barrier Lyapunov functions~\cite{ren2023vector,sturz2020distributed,zhang2022adaptive}. While these approaches explicitly encode stability and safety constraints, they typically rely on accurate system models. In contrast, model-free approaches~\cite{sun2019adaptive,zhang2022adaptive}, such as RL-based controllers~\cite{yang2022dynamic,liu2023reinforcement,zhou2024enhancing}, have demonstrated strong performance without requiring precise models. However, these methods generally lack formal guarantees of safety and stability. To the best of the authors’ knowledge, no existing work provides certified neural controllers for large-scale interconnected systems with cooperative safety and stability guarantees.

\noindent\textbf{Neural Certificates and Formal Verification.}
To provide guarantees for learning-based controllers, several works propose training with Lyapunov or barrier certificates~\cite{dai2021lyapunov,clark2021verification,qinlearning,yanglyapunov,yang2023model,zhang2023exact,zhang2023compositional,schlaginhaufen2021learning,pmlr-v270-hu25a,10886052,zhang2025gcbf+,mandal2024safe,nadali2024neural,zhou2025synthesis,neustroev2025neural,yu2025neural,10591251}. Such certificates can be formally verified with tools like $\alpha$-$\beta$-crown~\cite{wang2021beta}, Marabou~\cite{wu2024marabou}, and NNV~\cite{lopez2023nnv}. These tools further allow for counterexample-guided inductive synthesis (CEGIS)~\cite{ding2022novel,mandal2024formally,zhao2024neural} or certified training~\cite{mueller2023certified}, thereby systematically synthesizing controllers that satisfy barrier or Lyapunov properties. Nonetheless, to the best of the authors’ knowledge, no existing work has proposed neural certificates that account for cooperation and dynamic neighborhood structures in large-scale interconnected systems.

\section{Preliminary and Problem Statement}
\label{sec:Preliminary and Problem Statement}
In this section, we first introduce the preliminaries on Metzler and Hurwitz matrices in Section~\ref{subsec: Preliminary}, and then present the problem statement in Section~\ref{subsec: Problem Statement}.
\subsection{Preliminary}
\label{subsec: Preliminary}
\paragraph{Metzler matrix.}
A real square matrix \( K \in \mathbb{R}^{q \times q} \) is called a \emph{Metzler matrix} if all its off-diagonal entries are nonnegative:
\begin{equation}
K_{ij} \ge 0 \quad \text{for all } i \ne j.
\end{equation}
This structure naturally arises in interconnected or cooperative systems, where positive coupling between agents is common.

\paragraph{Hurwitz matrix.}
A real square matrix \( K \in \mathbb{R}^{q \times q} \) is called \emph{Hurwitz} if all its eigenvalues have strictly negative real parts:
\begin{equation}
\operatorname{Re}(\lambda^K_i) < 0 \quad \text{for all eigenvalues } \lambda_i \text{ of } K.
\end{equation}
This condition implies that the linear system \( \dot{z} = K z \) is globally exponentially stable at the origin.

\subsection{Problem Statement}
\label{subsec: Problem Statement}
Consider an interconnected system that consists of multiple agents in the set $\mathcal{N}$ with $q=|\mathcal{N}|$, where each agent is governed by its own dynamics and is associated with an open goal region $\mathbb{X}_{i,G}$ to be reached. Let $x_i \in \mathbb{X}_i\subset \mathbb{R}^{n}$ be the state of subsystem~$i$.  
Due to limited communication bandwidth, agents can only interact with at most $M_i-1\in\mathbb{N}$ neighbors that lie within a finite sensing radius $R_i^c>0$. Mathematically, let $\widetilde{\mathcal{N}}_i
:= \operatorname{Top}_{M_i-1}
\Bigl(
\left\{ j \in \mathcal{N}_i^c \setminus \{i\}
\;\middle|\;
\|x_i - x_j\| \le R_i^c
\right\}
\Bigr)$, where $\operatorname{Top}_{M_i-1}(\cdot)$ selects at most $M_i-1$ elements
with the smallest Euclidean distances $\|x_i-x_j\|$, $\mathcal N_i^c \subseteq \mathcal N$ denotes the set of agents that are
potentially communicable with agent~$i$, and hence the set $\widetilde{\mathcal{N}}_i$ represents the neighboring agents that can communicate with agent~$i$ and lie within its communication radius. We then define the local neighborhood as $\mathcal{N}_i = \widetilde{\mathcal{N}}_i \cup \{i\}$. Next, for each agent $i$, we define $\bar{x}_i\in\bar{\mathbb{X}}_{i}\subset\mathbb{R}^{M_i\times n}$ as the concatenated vector of $x_i$ and the states of neighbor nodes (i.e., $x_j,~j\in\widetilde{\mathcal{N}}_i$). Note that $\bar{x}_i$ is padded with a constant vector to a fixed size $M_i$ if $|\mathcal{N}_i|<M_i$.

The state of agent $i$ updates according to the following system dynamics:
\begin{equation}\label{eq: system dynamics}
    \dot{x}_i \;=\;
    f_i\!\bigl(\bar{x}_i\bigr)
    \;+\;
    g_i(\bar{x}_i)\,u_i,
\end{equation}
where $f_i: \mathbb{R}^{M_i\times n}\!\to\!\mathbb{R}^{n}$ and $g_i:\mathbb{R}^{M_i\times n}\!\to\!\mathbb{R}^{n\times m}$ are locally Lipschitz, and $u_i \in \mathbb{U}_i \subseteq \mathbb{R}^m$ is the control input within  a compact control limitation set $\mathbb{U}_i$. 

We aim to design decentralized control policies, denoted by $u_i=\pi_i\bigl(\bar{x}_i\bigr)$ for each agent $i\in\mathcal{N}$, that guarantee both \emph{safety} (avoiding unsafe regions) and \emph{liveness} (reaching goal regions) for each agent. This leads to the formulation of a multi-agent reach–while–avoid (MARWA) task, formally defined in Def.~\ref{def:CoRWA}.

\begin{definition}[Multi-Agent Reach–While–Avoid Task]
\label{def:CoRWA}

\noindent
Let the joint state of a team of $q$ agents be 
$\bm{x} = (x_1, \dots, x_q) \in \prod_{i=1}^q\mathbb{X}_i \subset \mathbb{R}^{q\times n}$. Each agent $i$ is associated with three given pairwise–disjoint open sets  
$\mathbb{X}_{i,I}$ (initial), $\mathbb{X}_{i,G}$ (goal), and $\mathbb{X}_{i,U}$ (unsafe), all contained in~$\mathbb{X}_i$. 

\vspace{3pt}\noindent
A collection of distributed control policies
$\bm{u} = (u_1, \dots, u_q)$ solves the MARWA task if every closed-loop trajectory for each agent~$i$ starting from initial set $X_{i,I}$ satisfies:

\begin{enumerate}[label=(\roman*),leftmargin=12pt]
  \item \textbf{Safety:}\;
        $x_{i}(t) \notin \mathbb{X}_{i,U},~\forall t \ge 0$, i.e., each agent does not enter its unsafe set;
  \item \textbf{Liveness:}\; 
Each agent $i$ eventually reaches and stays within its goal set, \begin{equation}
    \exists T_{i,G} \ge 0, \forall t\geq T_{i,G},\quad x_i(t) \in \mathbb{X}_{i,G}
\end{equation}
\end{enumerate}
\end{definition}

Typically, scalar Lyapunov and barrier functions~\cite{mandal2024formally,zhang2025gcbf+} are used to enforce the liveness and safety conditions of the MARWA task defined in Def.~\ref{def:CoRWA}. 
However, scalar Lyapunov and barrier functions require an input dimension equal to the sum of all agents' input dimensions, making their synthesis intractable for large-scale interconnected systems.
To address this limitation, we then introduce Dynamic-Localized Vector Control Lyapunov and Barrier Functions (DL-VCLFs and DL-VCBFs) to enhance tractability.

\section{Neural Cooperative Reach-While-Avoid Certificates}
\label{sec:Neural Cooperative Reach-While-Avoid Certificates}
To enhance the tractability of scalar Lyapunov and barrier functions, we propose DL-VCLFs and DL-VCBFs that explicitly leverage the agents' cooperative ability.  The proposed DL-VCLFs and DL-VCBFs are then formulated as neural cooperative reach-while-avoid (Co-RWA) certificates. By leveraging state-dependent local interactions, these vector-valued certificates enable more effective training and formal verification for decentralized control in interconnected systems.

To characterize agents' ability to cooperate, we define a state-dependent interaction mask that captures the local neighborhood structure of agent $i$ as follows:
\begin{equation}
A_{ij}(\bm{x})=\left\{
\begin{aligned}
    &1 ,\quad j\in\mathcal N_i(\bm{x})\\
    &0 ,\quad j\notin\mathcal N_i(\bm{x})\\
\end{aligned}\right.
\label{eq:mask}
\end{equation}
which indicates whether agent $j$ is considered a current neighbor of agent $i$. We further define $A_i(\bm{x}) = [A_{i1}(\bm{x}), A_{i2}(\bm{x}), \dots, A_{iq}(\bm{x})]^\top \in \{0,1\}^q$ as the neighborhood indicator vector of agent $i$, which will be used to localize the influence of neighboring agents in subsequent Lyapunov and barrier function constructions.

We first address the liveness of the MARWA problem by analyzing a sufficient condition, i.e., global asymptotic stability (GAS), under decentralized control while explicitly accounting for cooperation. Here, GAS ensures that agents are attracted to a pre-defined equilibrium point contained in the open goal set $\mathbb{X}_{i,G}$, which serves as a sufficient condition for liveness. We give the following definition and remark for detailed explanation.

\begin{definition}[Global asymptotic stability]
\label{def:GAS}
A system is said to be GAS if the equilibrium point $x^*$ is both Lyapunov stable and globally attractive~\cite{khalil2002nonlinear}. That is,
\begin{enumerate}
    \item For any $\varepsilon > 0$, there exists $\delta > 0$ such that $\|x(0) - x^*\| < \delta$ implies $\|x(t) - x^*\| < \varepsilon$ for all $t \ge 0$ (stability);
    \item For any $x(0) \in \mathbb{R}^n$, $\lim_{t \to \infty} x(t) = x^*$ (global attractivity).
\end{enumerate}
\end{definition}
\begin{remark}
The use of GAS in Def.~\ref{def:GAS} to guarantee liveness has been studied in~\cite{tan2021deductive} from a perspective of formal methods. Here, we reinterpret this implication from a control-theoretic viewpoint. 
For each agent $i\in \mathcal{N}$, let us assume that the equilibrium $x_i^*$ belongs to the goal set $\mathbb{X}_{i,G}$. Since $\mathbb{X}_{i,G}$ is an open set, there exists $\varepsilon_i > 0$ such that the open ball $\mathcal{B}_{\varepsilon_i}(x_i^*) = \{x \mid \|x-x_i^*\| < \varepsilon_i\}\subseteq \mathbb{X}_{i,G}$. Since $x_i^*$ is globally asymptotically stable, there exists
$\delta_i \in (0,\varepsilon_i)$ such that $\|x_i(0)-x_i^*\|<\delta_i$ implies $\|x_i(t)-x_i^*\|<\varepsilon_i$ for all $t \ge 0$. Moreover, global attractivity ensures that the trajectory enters $\mathcal{B}_{\delta_i}(x_i^*)$ (i.e., a $\delta_i$-neighborhood of the equilibrium) at some finite time $T_{i,\delta} \ge 0$. Consequently, for each agent $i$, $x_i(t)\in \mathcal{B}_{\varepsilon_i}(x_i^*) \subseteq \mathbb{X}_{i,G}$ for all $t \ge T_{i,\delta}$, which directly enforces the liveness property.
\end{remark}

To achieve the GAS while considering cooperation, we investigate a stronger notion, i.e., global exponential stability (GES). GES can be guaranteed under a sufficient condition, which extends VCLFs \cite{ren2023vector} to DL-VCLFs by considering dynamic neighborhoods. DL-VCLF is formalized in Def.~\ref{def:DL-VCLFs}. 

\begin{definition}[DL-VCLF]\label{def:DL-VCLFs}
For interconnected system~\eqref{eq: system dynamics}, a continuously differentiable map
$V:\mathbb{R}^{q\times n}\!\to\!\mathbb R_{\ge0}^{q}$ in the form $V(\bm{x}):=(V_1(x_1),\dots,V_q(x_q))$ is a dynamic-localized vector control Lyapunov function (DL-VCLF) if
\begin{itemize}
\item Each $V_i$ is positive–definite and radially unbounded, i.e., $V_i(0) = 0,\quad V_i(x_i) > 0, ~\text{for all } x_i \neq 0$, 
and $\lim_{\|x_i\|\to\infty} V_i(x_i) = \infty$.
\item There exists a vector $\lambda_i\in\mathbb{R}^q$ and $W_i(x) = \lambda_i\circ A_i(x)$, where $\circ$ is the element-wise product, such that:
\begin{equation}\label{eq:VCLF_condition}
\begin{aligned}
\inf_{u_i\in\mathbb U_i}
      \bigl\{L_{f_i}V_i\bigl(x_i\bigr)
        +L_{g_i}V_i(x_i)\pi_i\bigl(\bar{x}_i\bigr)\bigr\}\le W_i^{\top}(x)V(x).
\end{aligned}
\end{equation}
\item The matrix $\Lambda := (\lambda_1,\ldots,\lambda_q)\in\mathbb{R}^{q\times q}$ is Metzler and Hurwitz.
\end{itemize}
\end{definition}

Theorem~\ref{thm:dl-vclf-stab} establishes a DL-VCLF-based sufficient condition for GES of large-scale interconnected systems with dynamic, state-dependent neighborhoods. Since GES is a stronger notion than GAS, this result also guarantees GAS of the equilibrium. A detailed proof is provided in Appendix~\ref{app: theorem 1}.
\begin{theorem}[GES with DL-VCLF]
\label{thm:dl-vclf-stab}
Suppose $V$ is a DL-VCLF satisfying the conditions in Def.~\ref{def:DL-VCLFs}. If for each $i\in\mathcal{N}$, there exists locally Lipschitz continuous control
input $\bm{u} := (u_1, \dots, u_q) \in \mathbb{R}^{q\times m}$ such that 
the resulting closed-loop trajectories of the interconnected system  \eqref{eq: system dynamics}  satisfy \eqref{eq:VCLF_condition}, then the system \eqref{eq: system dynamics} is globally exponentially stable.
\end{theorem}

Second, we address the safety of the MARWA problem by constructing a DL-VCBF framework to certify safety in interconnected systems. The general process mirrors the construction of DL-VCLF. We construct the safe set in a compositional manner, in which the safe set for each agent is defined as follows:
\begin{align}
\mathcal{C}_i
&:= \Bigl\{\;\bar{x}_i \in \mathbb{R}^{M_i\times n} \;\Bigm|\;
          h_i(\bar{x}_i) \,\ge\, 0 \Bigr\},\label{eq:C_i}
\end{align}
Denote the boundary of set $\mathcal{C}_i$ as $\partial\mathcal{C}_i
:= \{\bar{x}_i \in \mathbb{R}^{M_i\times n} \mid
          h_i(\bar{x}_i) = 0 \}$ and the interior as 
$\operatorname{Int}(\mathcal{C}_i)
:= \{\bar{x}_i \in \mathbb{R}^{M_i\times n} \mid
          h_i(\bar{x}_i) > 0 \}$. 
Assume set $\mathcal{C}_i$ is non-empty and has no isolated
points, i.e., $\operatorname{Int}(\mathcal{C}_i)\neq\varnothing$ and the closure $\overline{\operatorname{Int}(\mathcal{C}_i)}=\mathcal{C}_i$. 
We can then compositionally define the safe set $\mathcal{C}:=\prod_{i=1}^{q}\mathcal{C}_i$ with $
\operatorname{Int}(\mathcal{C}):=\prod_{i=1}^{q}\operatorname{Int}(\mathcal{C}_i)$.

Then, DL-VCBF is formally defined as follows. 
\begin{definition}[DL-VCBF]\label{def:DL-VCBFs}
For interconnected system~\eqref{eq: system dynamics}, denote the joint extended state as $\bm{\bar{x}}=(\bar{x}_1,\cdots,\bar{x}_q)\in \bar{\mathbb{X}}^q\subset\mathbb{R}^{\sum_{i\in\mathcal{N}}M_i\times n}$. 
A continuously differentiable map $h:\mathbb{R}^{\sum_{i\in\mathcal{N}}M_i\times n}\to\mathbb{R}^{q}$ in the form $h(\bm{\bar{x}})=\bigl(h_1(\bar{x}_1),\dots,h_q(\bar{x}_q)\bigr)$ is said to be a dynamic-localized vector control barrier function (DL-VCBF) if for all $i\in\mathcal{N}$, 
\begin{equation}
\sup_{u_i \in \mathbb{U}_i}
\bigl\{L_{f_i} h_i(\bar{x}_i) + L_{g_i} h_i(\bar{x}_i)\pi_i\bigl(\bar{x}_i\bigr) \bigr\}
\ge
\Gamma_i(\bm{x})^{\top} h(\bm{x}),
\label{eq: CBF}
\end{equation}
where $\Gamma_i(\bm{x})=\mu_i\circ A_i(\bm{x})$, $\Upsilon := (\mu_1,\dots,\mu_q) \in \mathbb{R}^{q\times q}$ is Metzler.
\end{definition}

Theorem~\ref{thm:dl-vblf-safe} shows that the existence of DL-VCBF suffices for safety guarantees. Specifically, by ensuring that the Lie derivative of each local barrier function exceeds a cooperative lower bound, the overall system can maintain forward invariance of the safe set, i.e., trajectories starting in the safe set never leave it, which is our notion of safety. The detailed proof is provided in Appendix~\ref{app: theorem 3}.
\begin{theorem}[Forward-invariant safe set with DL-VCBF]\label{thm:dl-vblf-safe}
Consider the safe set $\mathcal{C} = \prod_{i=1}^{q} \mathcal{C}_i$ with $\mathcal{C}_i$ defined in \eqref{eq:C_i}.  
Suppose the system~\eqref{eq: system dynamics} admit a DL-VCBF  $h : \mathbb{R}^{\sum_{i\in\mathcal{N}}M_i n} \rightarrow \mathbb{R}^q.$
If for each $i \in \mathcal{N}$, there exists a Lipschitz continuous function $u_i$ satisfying \eqref{eq: CBF}, then the controller $\bm{u} := (u_1, \dots, u_q)$ renders the set $\mathcal{C}$ forward invariant.
\end{theorem}

Building upon these theoretical foundations, we define the Co-RWA certificate to effectively achieve MARWA as follows:

\begin{definition}[Cooperative Reach-While-Avoid Certificate]
\label{def:Co-RWA}
\noindent We define the \emph{Co-RWA certificate} for an interconnected system as a collection of DL-VCLFs and DL-VCBFs that jointly ensure global exponential stability and safety under decentralized control. Specifically, a set of functions $\{V_i, h_i\}_{i \in \mathcal{N}}$ constitutes a Co-RWA certificate if:
\begin{itemize}
\item Each $V_i$ satisfies the DL-VCLF condition in Def.~\ref{def:DL-VCLFs}, guaranteeing exponential convergence to goal regions;
\item Each $h_i$ satisfies the DL-VCBF condition in Def.~\ref{def:DL-VCBFs}, ensuring forward invariance of the corresponding safe set $\mathcal{C}_i$.
\end{itemize}
\end{definition}
Such a certificate provides a decentralized and compositional guarantee that all agents in the system will cooperatively reach their goals while avoiding unsafe regions, which can enforce the task described in Def.~\ref{def:CoRWA}.

We next present a training and verification framework that synthesizes neural certificates and controllers with formal guarantees of liveness and safety.

\section{Framework for Synthesizing and Verifying Cooperative RWA Certificates}
\label{sec:framework}
\begin{figure*}[t]
    \centering
    \includegraphics[width=0.8\textwidth]{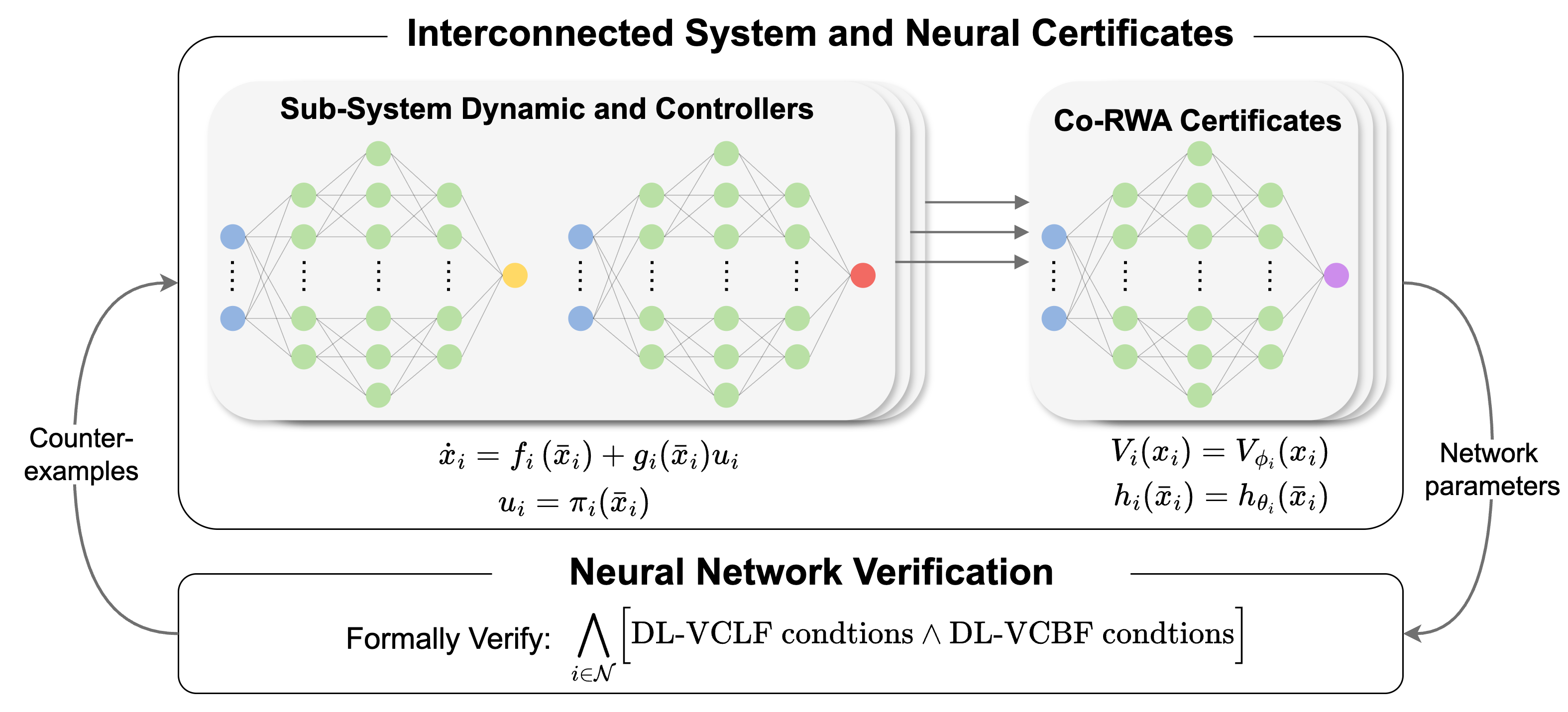}
    \caption{Overview of the proposed framework for controller synthesis and neural certificate verification in interconnected systems.}
    \label{fig:framework}
\end{figure*}
This section presents a unified framework for synthesizing and verifying neural Co-RWA certificates and controllers that ensure both safety and liveness as in Figure~\ref{fig:framework}. Section~\ref{subsec: controller synthesis} introduces the training formulation for neural network-based controllers and cooperative certificates. Section~\ref{subsec: verification formulation} details the verification formulation considering the discretization and approximation error. Scalability analysis is discussed in Section~\ref{subsec: Scalability Analysis}.

\subsection{Control Policy Synthesis}
\label{subsec: controller synthesis}
We now develop a learning-based framework to synthesize decentralized control policies with global exponential stability and safety guarantees. The core idea is to jointly learn Lyapunov and barrier certificates along with a policy. This enables the integration of formal guarantees into policy learning, ensuring that the resulting controllers have certifiable safety and global exponential stability.
Specifically, we aim to simultaneously learn a neural DL-VCLF $V_{\bm{\phi}}(\bm{x}) = [V_{\phi_i}(x_1), \dots, V_{\phi_q}(x_q)]^\top$ and a neural DL-VCBF $h_{\bm{\theta}}(\bm{x}) = [ h_{\theta_1}(\bar{x}_i), \dots, h_{\theta_q}(\bar{x}_q)]^\top$, while optimizing controllers $\pi(\bm{x}) = [\pi_1(\bar{x}_i), \dots, \pi_q(\bar{x}_q)]^\top$.

The learning objective is formalized as the following optimization problem:
\begin{align}
\min_{\pi, V_{\bm{\phi}}, h_{\bm{\theta}}}~ &
\sum_{i \in \mathcal{N}} \| \pi_i(\bar{x}_{i}) - \pi_{i,\text{nom}}(\bar{x}_{i}) \|^2
\label{eq:ft_obj}
\\
\text{s.t.}~~ &
L_{f_i} V_{\phi_i}(x_i) + L_{g_i} V_{\phi_i}(x_i)\, \pi_i(\bar{x}_i)
\le W_i^\top V_{\bm{\phi}}(\bm{x}),\notag\\
&\forall x_i \in \mathbb{X}_i,\forall \bar{x}_i \in\mathbb{\bar{X}}_i,\bm{x}\in \mathbb{X}^q,~ i \in \mathcal{N}, \label{eq:ft_lyap_constraint}
\\
& L_{f_i} h_{\theta_i}(\bar{x}_i) + L_{g_i} h_{\theta_i}(\bar{x}_i)\, \pi_i(\bar{x}_i)
\ge \Gamma_i^\top h_{\bm{\theta}}(\bm{x}),\notag\\
&\forall \bar{x}_i \in \mathbb{\bar{X}}_i,\bm{x}\in \mathbb{X}^q,~ i \in \mathcal{N},\label{eq:ft_barrieR_i^constraint}
\\
& V_{\phi_i}(x_i) > 0, \quad\forall x_i \ne 0,~ x_i \in \mathbb{X}_i,\label{eq:lyap_positive}\\
& h_{\theta_i}(\bar{x}_i) \geq \epsilon_0, \quad\forall \bar{x}_i \in \mathbb{X}_{i}\setminus\mathbb{X}_{i,U},\label{eq:barrier_positive}\\
& h_{\theta_i}(\bar{x}_i) < 0, \quad\forall \bar{x}_i \in \mathbb{X}_{i,U},\label{eq:barrier_negative}
\end{align}
where $\pi_{i,\text{nom}}$ is a nominal controller (e.g., obtained from reinforcement learning or imitation learning). $\epsilon_0$ is a small positive constant that enforces a strict safety margin for the barrier function.

Since the optimization problem cannot be solved in closed form, we reformulate it using the following loss function:
\begin{align}
&\mathcal L=
\sigma_1L_{\text{ctrl}}
+\sigma_2L_{\text{DL-VCLF}}
+\sigma_3L_{\text{DL-VCBF}},\label{eq: loss}\\
&L_{\text{ctrl}} =
 \frac{1}{|\mathcal{N}|}\sum_{i\in\mathcal{N}}\|\pi_i(\bar{x}_i)-\pi_{i,\text{nom}}(\bar{x}_i)\|^2,\\
&L_{\text{DL-VCLF}} =\frac{1}{|\mathcal{N}|}\left(\sum_{i\in\mathcal{N}}
 \operatorname{ReLU}\bigl(
   L_{f_i}V_{\phi_i}+L_{g_i}V_{\phi_i}\,\pi_i\right.\\
&\left.-W_i^{\top}(\bm{x}) V_{\bm{\phi}}+\varepsilon_1\bigr)+\sum_{i\in\mathcal{N}} \operatorname{ReLU}\left(V_{\phi_i}+\varepsilon_2\right)\right)\notag\\
&L_{\text{DL-VCBF}} =
 \frac{1}{|\mathcal{N}|}\left(\sum_{i\in\mathcal{N}}
 \operatorname{ReLU}\bigl(
   \Gamma_i^{\top}(\bm{x}) h_{\bm{\theta}}
   -L_{f_i}h_{\theta_i}\right.\\
&-L_{g_i}h_{\theta_i}\,\pi_i+\varepsilon_3\bigr)+\sum_{i\in\mathcal{N},\bar{x}_i\in\mathbb{X}_{i,U}}\operatorname{ReLU}(h_{\theta_i}+\epsilon_0+\varepsilon_4)+\notag\\
&\left.\sum_{i\in\mathcal{N},\bar{x}_i\in\mathbb{X}_{i}\setminus\mathbb{X}_{i,U}}
 \operatorname{ReLU}\bigl(-h_{\theta_i}+\varepsilon_5\bigr)\right)\notag
\end{align}
where $\sigma_{1},\sigma_{2},\sigma_{3}>0$ are weighting coefficients that balance imitation accuracy, Lyapunov feasibility, and barrier feasibility. $\operatorname{ReLU}(a)=\max\{a,0\}$ is the usual hinge penalty, so each constraint only contributes to the loss when it is violated, $\varepsilon_{1},\dots,\varepsilon_{5}>0$ are small slack variables added for numerical robustness. In practice, since dynamical systems are typically simulated in discrete time, a common approach is to perform a one-step simulation $x_i \rightarrow x_i^{\text{next}}$ and estimate $\left[\nabla h_i(\bar{x}_i)\right]^\top \left(
    f_i\!\bigl(\bar{x}_i\bigr)
    \;+\;
    g_i(\bar{x}_i)\,u_i\right)$ by using $\left(h_i(\bar{x}_i^{\text{next}}) - h_i(\bar{x}_i)\right) / T$, where $T$ denotes the sampling interval.
    
\subsection{Verification Formulation}
\label{subsec: verification formulation}

To verify whether the synthesized controller satisfies Co-RWA, we first discretize the continuous-time dynamics, as neural network verification tools and mixed-integer optimization-based methods are inherently designed for discrete-time systems. We then revise the Co-RWA conditions to explicitly account for discretization errors. Finally, we formulate the verification task as a neural network verification problem and accordingly leverage the resulting counterexamples to iteratively refine the learned neural controllers and certificates. 

Specifically, we discretize the continuous-time dynamics using a forward Euler scheme with sampling interval $T$, and learn neural network models $\tilde f_i, \tilde g_i$ from sampled data.
\begin{align}
\label{eq: discrete_system}
   &\hat{x}_i((k+1)T)
   = x_i(kT) + T\Bigl(
   \tilde{f}_i\bigl(\bar{x}_i(kT)\bigr)\notag\\
   &+ \tilde{g}_i\bigl(\bar{x}_i(kT)\bigr)u_i(kT)\Bigr).
\end{align}

This introduces two sources of mismatch:
(1) \emph{discretization error} between the continuous-time trajectories
and the Euler scheme used in verification; and (2) \emph{approximation error} between the true dynamics and their neural representations. We next derive uniform bounds on both sources of mismatch and accordingly propagate these bounds to the Lyapunov (and barrier) inequalities so that conclusions drawn from the neural discrete-time model remain valid for the true continuous-time system. 

To this end, we first make Assumptions~\ref{asmp: Lipschitz} and~\ref{asmp: uniform sample} on the Lipschitz continuity of relevant functions and the error bounds of neural network approximations $\tilde f_i, \tilde g_i$. 

\begin{assumption}[Lipschitz continuity]
\label{asmp: Lipschitz}
Consider the interconnected system in~\eqref{eq: system dynamics}.

(i) For each $i\in\mathcal N$, the maps $\dot{x}_i=f_i(\bar{x}_i)+g_i(\bar{x}_i)u_i(\bar{x}_i)$
are globally Lipschitz in $\bar{x}_i(t)$ with Lipschitz constants
$L_{x_i}$.

(ii) There exist constants $M_{x_i}$ such that
\begin{align}
M_{x_i} := \sup_{\bar x_i,u_i}\bigl\|
f_i(\bar x_i)+g_i(\bar x_i)u_i\bigr\| < \infty
\end{align}
for all $\bar{x}_i$ in the operating domain $\bar{\mathbb{X}}_i$.

(iii) $V_i$ and its time derivative $\dot{V}_i$ along the trajectories of~\eqref{eq: system dynamics}
are Lipschitz continuous with Lipschitz constants $L_{V_i}$ and $L_{\dot{V}_i}$,
respectively, and satisfy $\|\dot{V}_i\|\le M_{V_i}$ on the operating domain.

(iv) $h_i$ and its time derivative $\dot{h}_i$ along the trajectories of~\eqref{eq: system dynamics}
are Lipschitz continuous with Lipschitz constants $L_{h_i}$ and $L_{\dot{h}_i}$,
respectively, and satisfy $\|\dot{h}_i\|\le M_{h_i}$ on the operating domain.
\end{assumption}

\begin{remark}\label{rmk:lipschitz}
In our setting, such Lipschitz continuity assumptions are mild and are automatically satisfied for a broad class of smooth physical systems on compact operating domains. The calculation of the Lipschitz constants in
Assumption~\ref{asmp: Lipschitz} are detailed as follows.  

\noindent(i) For each agent $i\in\mathcal N$, the constants $L_{x_i}$ are
derived a priori from the analytic model on the compact operating set and
treated as fixed.  

\noindent(ii) The Lyapunov and barrier functions $V_i$ and $h_i$ are implemented as
feed-forward neural networks with smooth Lipschitz activation functions (e.g.,
Softplus, $\tanh$). Hence $V_i,h_i$ and their derivatives are Lipschitz on the
prescribed domain. The corresponding Lipschitz constants (including those for
$\dot V_i,\dot h_i$) are estimated offline using a neural Lipschitz
bounding method, e.g.,~\cite{xu2024eclipse}.
\end{remark}

\begin{assumption}[Bounded Approximation Error]
\label{asmp: uniform sample}
Given a bounded domain $\mathcal{Z}_i \subset \mathbb{R}^{m}$ for agent $i$ with 
$m = n_i + \sum_{j\in\mathcal{E}_i} n_j$, let 
$z_{i,k} := (x_{i,k},\{x_{j,k}\}_{j\in\mathcal{E}_i})\in\mathcal{Z}_i$ 
collect the local state and neighbor states. 
Construct a dataset $\mathcal{D}_i \subset \mathcal{Z}_i$ by discretizing each coordinate of $\mathcal{Z}_i$ on a rectangular grid with per-dimension step sizes $\boldsymbol{\Delta}=(\Delta_1,\dots,\Delta_m)\in\mathbb{R}^m_{>0}$. 
Let $\tilde f_i$ and $\tilde g_i$ be surrogate models of $f_i$ and $g_i$, respectively, and denote the true and approximated closed-loop dynamics by
\begin{align}
&F_i(z_{i,k},u_{i,k}) := f_i(z_{i,k}) + g_i(z_{i,k}) u_{i,k},\\
&\tilde F_i(z_{i,k},u_{i,k}) := \tilde f_i(z_{i,k}) + \tilde g_i(z_{i,k}) u_{i,k}.
\end{align}

Let $\hat{\epsilon}_i$ denote the empirical maximum approximation error of the surrogate closed-loop dynamics on $\mathcal{D}_i$, written as
\begin{align}
\hat{\epsilon}_i \;:=\; \max_{(z_{i,k},u_{i,k})\in \mathcal{D}_i} 
\, \bigl| F_i(z_{i,k},u_{i,k}) - \tilde F_i(z_{i,k},u_{i,k}) \bigr|.
\end{align}
We assume $\hat{\epsilon}_i<+\infty$, meaning that the approximation error is bounded on the discretized grid $\mathcal{D}_i$. 
\end{assumption}

Under the above assumptions, both (i) the one-step discretization error between the continuous-time and sampled dynamics and (ii) the approximation error introduced by the neural network surrogates can be uniformly bounded on the operating domain. These bounds allow us to rigorously relate the decrement conditions verified on the learned discrete-time model~\eqref{eq: discrete_system} to those of the underlying continuous-time interconnected system~\eqref{eq: system dynamics}, as stated in the following theorem.

\begin{theorem}[Sufficient Conditions for Co-RWA via Discrete Time Verification]
\label{thm:disc-CoRWA}
Consider the interconnected system~\eqref{eq: system dynamics} under
Assumptions~\ref{asmp: Lipschitz} and~\ref{asmp: uniform sample}. 

Suppose the following conditions are satisfied:

(i) Each $V_i$ is positive-definite and radially unbounded.

(ii) There exists a sampling period $T>0$ and Metzler matrices
$\Lambda=(\lambda_1,\dots,\lambda_q)$ and
$\Upsilon=(\mu_1,\dots,\mu_q)$ with $\Lambda$ Hurwitz such that, for every
$k\in\mathbb N$ and every $i\in\mathcal N$, the Euler approximation of the Lyapunov condition satisfies
\begin{align}
\frac{V_i\bigl(\hat x_i((k+1)T)\bigr)-V_i\bigl(x_i(kT)\bigr)}{T}
&\le \lambda_i^\top V\bigl(x(kT)\bigr) - e^{V}_{i,\sup}, \label{eq:disc-V-CoRWA}\\
\frac{h_i\bigl(\hat{\bar x}_i((k+1)T)\bigr)-h_i\bigl(\bar x_i(kT)\bigr)}{T}
&\ge \mu_i^\top h\bigl(x(kT)\bigr) + e^{h}_{i,\sup}, \label{eq:disc-h-CoRWA}
\end{align}
where
\begin{align}
e^{V}_{i,\sup} &:= \tfrac{1}{2}T\bigl(L_{V_i}L_{x_i} + L_{\dot V_i}\bigr)M_{x_i} + L_{V_i}\hat{\epsilon}_i,\\
e^{h}_{i,\sup} &:= \tfrac{1}{2}T\bigl(L_{h_i}L_{x_i} + L_{\dot h_i}\bigr)M_{x_i} + L_{h_i}\hat{\epsilon}_i,
\end{align}
with $L_{\dot V_i}$ and $L_{\dot h_i}$ representing the Lipschitz constants of
$\dot V_i$ and $\dot h_i$, and $L_{V_i}$, $L_{h_i}$ denoting the Lipschitz constants of
$V_i$ and $h_i$, respectively.

Then, the corresponding continuous-time Co-RWA matrix inequalities
\begin{align}
\dot V_i(x) \le \lambda_i^\top V(x),\qquad
\dot h_i(\bar x_i) \ge \mu_i^\top h(x),
\quad \forall i\in\mathcal N,
\end{align}
hold on the operating domain. Consequently,
$\{V_i,h_i\}_{i\in\mathcal N}$ constitute a neural Co-RWA certificate in the sense of Def.~\ref{def:CoRWA}.
\end{theorem}

Based on Theorem~\ref{thm:disc-CoRWA}, the verification query can be formulated as follows:
\begin{equation}
    \bigwedge_{i \in \mathcal{N}}\Bigl[ \eqref{eq:lyap_positive}\land\eqref{eq:barrier_positive}\land\eqref{eq:barrier_negative}\land\eqref{eq:disc-V-CoRWA}\land\eqref{eq:disc-h-CoRWA}\Bigr]
    \label{eq: verification for all agents}
\end{equation}
To accelerate verification, we adopt this distributed formulation as in Eq.~\eqref{eq: verification for all agents}, which decomposes a large verification query on a complex multi-agent system into a number of smaller verification queries, each involving the analysis and encoding of a single agent’s controller. 

With the training and verification formulation, we use a counterexample-guided inductive synthesis (CEGIS) loop to obtain fully verified controllers and certificates as in Algorithm~\ref{alg:joint-training-cegis}. At each CEGIS iteration, we jointly train $\{V_{\phi_i}\}_{i\in\mathcal{N}}$, $\{h_{\theta_i}\}_{i\in\mathcal{N}}$ and, $\{\pi_i\}_{i\in\mathcal{N}}$ until the loss Eq.~\eqref{eq: loss} converges and then use a formal neural network verifier (e.g., \cite{nnvTwo,wu2024marabou,wang2021beta,xu2020automatic,verinet}) to verify the certificate. If the verifier identifies counterexamples violating Eq.~\eqref{eq: verification for all agents}, to accelerate the CEGIS loop, we sample points in the proximity of the counterexample and use these to augment the training data. This process is repeated iteratively until no counterexamples are found.

\begin{algorithm}[h]
\caption{Counterexample-Guided Inductive Synthesis Loop for Co-RWA Certificate}
\label{alg:joint-training-cegis}
\begin{algorithmic}[1]
\Require Initial controller parameters $\{\pi_i\}_{i\in\mathcal{N}}$, initial dataset $\mathcal{D}$ containing state-action pairs, Lyapunov and Barrier function $\{V_{\phi_i}\}_{i\in\mathcal{N}}, \{h_{\theta_i}\}_{i\in\mathcal{N}}$, and a neural network verifier.
\Ensure Verified controllers $\{\pi_i\}_{i\in\mathcal{N}}$, Lyapunov and Barrier functions $\{V_{\phi_i}\}_{i\in\mathcal{N}},\{h_{\theta_i}\}_{i\in\mathcal{N}}$.
\State \textbf{Initialize:} Load controller parameters, initialize Lyapunov and Barrier function.
\Repeat
    \State Train $\{\pi_i\}_{i\in\mathcal{N}}$, $\{V_{\phi_i}\}_{i\in\mathcal{N}}$, and $\{h_{\theta_i}\}_{i\in\mathcal{N}}$ by minimizing the loss function~Eq.~(\ref{eq: loss}).
    \State Verify the Co-RWA conditions~Eq.~(\ref{eq:lyap_positive})-(\ref{eq:barrier_negative}),\eqref{eq:disc-V-CoRWA},\eqref{eq:disc-h-CoRWA} using the neural network verifier.
    \If{counterexamples violating the constraints are found}
        \State Augment dataset $\mathcal{D}$ with counterexamples.
    \EndIf
\Until{no counterexamples are found after verification.}
\end{algorithmic}
\end{algorithm}

\subsection{Scalability Analysis}
\label{subsec: Scalability Analysis}
The proposed framework for synthesizing Co-RWA-guaranteed neural network-based controllers can be computationally challenging for large-scale interconnected systems, due to the NP-hardness of the optimization problem and the slow convergence rate of the counterexample-guided training process. To address this, we develop a scalable synthesis and verification framework that accelerates controller synthesis and verification by reusing the controllers and neural Co-RWA certificates for smaller or structurally equivalent systems.

\noindent\textbf{Notation}.  
We write an interconnected system as $\mathcal I=(\mathcal N,\{\mathcal N_i(\bm{x})\}_{i\in\mathcal N},\{f_i\}_{i\in\mathcal N}),\{g_i\}_{i\in\mathcal N})$.
A subsystem  
\(
   \mathcal I'=(\mathcal N',\{\mathcal N'_i(\bm{x})\}_{i\in\mathcal N'},\{f'_i\}_{i\in\mathcal N'},\{g_i'\}_{i\in\mathcal N})
   \subset\mathcal I
\)
implies:
(i) $\mathcal N'\subset\mathcal N$;  
(ii) $\mathcal N'_i(\bm{x})=\{(i,j)\in\mathcal N_i(\bm{x})\mid j\in\mathcal N'\}$, and (iii) $f_i'\bigl(\bar{x}_i\bigr)=f_{i}\bigl(\tilde{\bar{x}}_{i}\bigr), g_i'(\bar{x}_i)=g_i(\tilde{\bar{x}})$, $\forall i \in \mathcal{N'}$.

To enable scalable certificate reuse, we formalize the structural similarity between interconnected systems using the notion of substructure isomorphism. This concept captures when a smaller or simpler subsystem shares identical local dynamics and neighborhood topologies with parts of a larger system. Such structural equivalence allows direct transfer of controllers and certificates without retraining or re-verification. We begin by defining this relationship in the context of dynamic interaction graphs:
\begin{definition}[Substructure Isomorphism]\label{dfn:stru_eq_RWA}
An interconnected system  
$\widetilde{\mathcal I}
    =(\widetilde{\mathcal N},\{\widetilde{\mathcal N}_j(\bm{x})\}_{j\in\widetilde{\mathcal N}},\{\widetilde f_j\}_{j\in\widetilde{\mathcal N}},\{\widetilde g_j\}_{j\in\widetilde{\mathcal N}})$ is \emph{substructure–isomorphic} to  
$\mathcal I
    =(\mathcal N,\{\mathcal N_i(\bm{x})\}_{i\in\mathcal N},\{f_i\}_{i\in\mathcal N}),\{g_i\}_{i\in\mathcal N})$
if there exists an injective map
$\tau:\widetilde{\mathcal N}\!\to\!\mathcal N$ such that  
\begin{equation}
\begin{aligned}
&\tau\bigl(\widetilde{\mathcal N}_j(\bm{x})\bigr)
           =\mathcal N_{\tau(j)}(\bm{x}),
\widetilde f_j=f_{\tau(j)},\widetilde{g}_j=g_{\tau(j)}\\
&\forall j\in\widetilde{\mathcal N},\;
       \forall x\in \mathbb{X}^{q}.
\end{aligned}
\end{equation}
\end{definition}

This definition ensures that the subsystem $\widetilde{\mathcal{I}}$ forms a closed subgraph of $\mathcal{I}$ under the mapping $\tau$, preserving both neighborhood structure and dynamics internally. The notion of substructure isomorphism provides a principled basis for reusing learned controllers and neural certificates across structurally similar systems. The following theorem shows that if two systems are substructure-isomorphic, then the Co-RWA certificates, along with their corresponding controllers, can be directly transferred from one to the other.

\begin{theorem}\rm{\textbf{(Co-RWA Preservation under Substructure Isomorphism)}}
\label{thm:RWA-Equivalence}
Let  
\(
   \mathcal I=(\mathcal N,\{\mathcal N_i(\bm{x})\}_{i\in\mathcal{N}},\{f_i\}_{i\in\mathcal{N}})
\)
admit decentralized controllers $\{\pi_i\}_{i\in\mathcal N}$
and associated DL-VCLF and DL-VCBF certificates  
$\{V_i,h_i\}_{i\in\mathcal N}$
satisfying conditions~\eqref{eq:ft_lyap_constraint}–\eqref{eq:barrier_negative}.  
If another system  
\(
   \widetilde{\mathcal I}
     =(\widetilde{\mathcal N},\{\widetilde{\mathcal N}_j(\bm{x})\}_{j\in\widetilde{\mathcal{N}}},\{\widetilde f_j\}_{j\in\widetilde{\mathcal{N}}})
\)
is substructure-isomorphic to $\mathcal I$
in the sense of Def.~\ref{dfn:stru_eq_RWA},
then it inherits the decentralized controllers  
\(
   \widetilde\pi_j=\pi_{\tau(j)}
\)
and the certificates  
\(
   \widetilde V_j=V_{\tau(j)},\;
   \widetilde h_j=h_{\tau(j)}
   \;(j\in\widetilde{\mathcal N}),
\)
and the resulting closed-loop dynamics of
$\widetilde{\mathcal I}$ fulfil the Co-RWA specification of Def.~\ref{def:Co-RWA}.
\end{theorem}

This theorem provides the theoretical foundation for certificate and controller reuse across different interconnected systems, with the proof presented in Appendix 4. In practice, once DL-VCLF and DL-VCBF certificates are learned and verified for a smaller system $\widetilde{\mathcal{I}}$, they can be transferred to a larger or different system $\mathcal{I}$ as long as it is substructure-isomorphic to $\widetilde{\mathcal{I}}$. This eliminates the need for retraining or re-verification in each new deployment, significantly improving the scalability of our framework. 

\section{Numerical Simulation}
\label{sec:Numerical Simulation}

In this section, we conduct numerical simulations to evaluate the performance of the proposed control framework. Section~\ref{subsec: system dynamics} introduces the application environments of the proposed method, including omnidirectional robots~\cite{ren2023vector} and vehicle platoons~\cite{ZHOU2024104885}. Section~\ref{subsec: simulation setting} presents the detailed simulation settings. Section~\ref{subsec: controller results} presents the control performance for two scenarios. Section~\ref{subsec: verification results} introduces the verification effectiveness of the proposed method. Section~\ref{subsec: Ablation} shows the ablation study. Some additional results are presented in Appendix~\ref{app: additional results}.

\subsection{System Dynamics}
\label{subsec: system dynamics}
Next, we present the model-based dynamics for the two scenarios. Although these dynamics are derived from physical models, we fit neural networks to approximate them. The learned models are then used for training and formal verification. In practice, to ensure robustness despite modeling errors, we can incorporate empirical conservative error bounds into the verification process and adopt robust control formulations where necessary.

\noindent \textbf{Omnidirectional robots}: Consider a team of $q \in \mathbb{N}$ omnidirectional robots~\cite{ren2023vector}. The dynamics of each robot are:
\begin{equation}
\dot{x}_i = f_i(x) + g_i(x_i) u_i,
\end{equation}
where state of each robot is $x_i = (x_{i1}, x_{i2}, x_{i3}) \in \mathbb{R}^3$, with $p_i = (x_{i1}, x_{i2}) \in \mathbb{R}^2$ is the position and $x_{i3}$ is the orientation. $u_i = (u_{i1}, u_{i2}, u_{i3}) \in \mathbb{R}^3$ is the wheel velocity. The nominal controller is obtained via imitation learning of a potential controller. The interaction among robots is:
\begin{align}
&f_i(x) := \left( f_{i1}(x), f_{i2}(x), 0 \right), \\
&f_{il}(x) := \sum_{j \in \mathcal{N}_i,\, j \ne i} \frac{k_i(x_{il} - x_{jl})}{\|p_i - p_j\| + \varepsilon_i}, \quad l \in \{1,2\},
\end{align}
where $\mathcal{N}_i$ is the neighbor set of robot $i$ and $k_i > 0$, and $\varepsilon_i$ is a small positive constant for numerical stability. The input matrix is defined as:
\begin{align}
&g_i(x_i) =
\begin{bmatrix}
\cos(x_{i3}) & -\sin(x_{i3}) & 0 \\
\sin(x_{i3}) & \cos(x_{i3}) & 0 \\
0 & 0 & 1
\end{bmatrix}
(J_i^\top)^{-1} R_i, \\
&J_i =
\begin{bmatrix}
0 & \cos(\tfrac{\pi}{6}) & -\cos(\tfrac{\pi}{6}) \\
-1 & \sin(\tfrac{\pi}{6}) & \sin(\tfrac{\pi}{6}) \\
L_i & L_i & L_i
\end{bmatrix}.
\end{align}

\noindent \textbf{Vehicle platoon}: Consider a vehicle platoon with one leading vehicle and $q$ following vehicles. The state of each vehicle is $x_i=(s_i,v_i)$, where $s_i(t)$ and $v_i(t)$ represent the spacing and velocity of vehicle $i$, respectively.

Each following vehicle's dynamics is described by second-order ordinary differential equations: 
\begin{align}
    &\dot{s}_{i}(t) =v_{i-1}(t)-v_{i}(t), \label{eq:s}\\
    &\dot{v}_{i}(t) = u_i(t), 
     \label{eq:v}
\end{align}
where the control policy of the following vehicle $u_i(t)$ is derived using reinforcement learning, as proposed in~\cite{ZHOU2024104885}. 

\subsection{Simulation Settings}
\label{subsec: simulation setting}
\noindent\textbf{Hardware and System Configuration.}
All experiments were conducted on a Linux machine (Ubuntu 24.04) equipped with an Intel Core Ultra 9 285K processor (24 cores), 125 GB of RAM, and dual NVIDIA GeForce RTX 5090 GPUs (each with 32 GB memory). The system runs Linux kernel 6.14.0 with CUDA version 12.8 and NVIDIA driver 570.153.02.

\noindent\textbf{Multi-Robot Simulation Setup.}
We consider a team of four omnidirectional robots, including one leader and three followers, performing formation tracking and obstacle avoidance. The simulation time step is set to $\Delta t = 0.05$,s with a total of $600$ steps (30 s total duration). Each robot is equipped with wheels of radius $r = 0.02$ m and a wheel-to-center distance $L = 0.2$ m. The wheel angular velocity is bounded by $|u| \leq 40$ rad/s. The maximum translational velocity is clipped at $1.0$ m/s for safety. Each agent senses its surroundings within a radius of $2.0$ m. The leader is assigned a global target position at $\boldsymbol{p}_1^* = [20.0, 0.0]^\top$ m. The followers are expected to maintain a triangular formation with respect to the leader, specified by offset vectors: follower 1 at $[-1.0, 1.0]^\top$ m, follower 2 at $[-1.0, -1.0]^\top$ m, and follower 3 at $[-2.0, 0.0]^\top$ m relative to the leader. To evaluate obstacle avoidance capability, three static circular obstacles are placed in the workspace:
\begin{itemize}
\item Obstacle 1 centered at $[6.0, 1.0]^\top$ m with radius $1.0$ m,
\item Obstacle 2 centered at $[10.0, -1.5]^\top$ m with radius $1.2$ m,
\item Obstacle 3 centered at $[14.0, 1.0]^\top$ m with radius $1.1$ m.
\end{itemize}

The nominal controllers of the agents are artificial potential fields: attractive fields with gains $k_{\text{target}} = 0.8$ for target tracking and $k_{\text{form}} = 0.8$ for formation maintenance; repulsive fields with $k_{\text{obs}} = 6.0$ for obstacle avoidance and $k_{\text{agent}} = 1.2$ for inter-agent separation. These fields have influence radii $d_{\text{obs}} = 2.5$,m and $d_{\text{agent}} = 0.1$,m. Additionally, a spring-coupling term with gain $k_{\text{cpl}} = 0.1$ is applied when agent separation falls below $\varepsilon_{\text{cpl}} = 0.1$,m.

\noindent\textbf{Vehicle Platoons Simulation Setup.} We train a vehicle platoon composed of one leading vehicle and three following connected and autonomous vehicles (CAVs), indexed as vehicles $0$–$3$. The leading vehicle follows either stochastic velocity perturbations or real-world driving trajectories (e.g., from Waymo~\cite{hu2022processing}). All vehicles share their states (e.g., positions, velocities) through communication, enabling full-information feedback for each CAV controller. The simulation proceeds in discrete time with step size $\Delta t = 0.1\,\mathrm{s}$ for a maximum of $2000$ steps. Each vehicle follows a longitudinal kinematic model with control inputs bounded in $[-5, 5]\,\mathrm{m/s}^2$.

\noindent\textbf{CEGIS Loop Setup.} Each CEGIS loop is trained for a maximum of 50 epochs, with a maximum of 100 iterations. The learning rate is initialized at 0.001 and follows a decay schedule. The initial dataset comprises 30,000 state pairs, with 80\% used for training and 20\% for validation, and a batch size of 32. To augment the counter-example set, Gaussian noise is applied to each original counter-example, generating 20 variants per instance.

\subsection{Control Performance Analysis}
\label{subsec: controller results}

\begin{figure}[t]
  \centering
  \begin{subfigure}[b]{\linewidth}
    \centering
    \includegraphics[width=\linewidth]{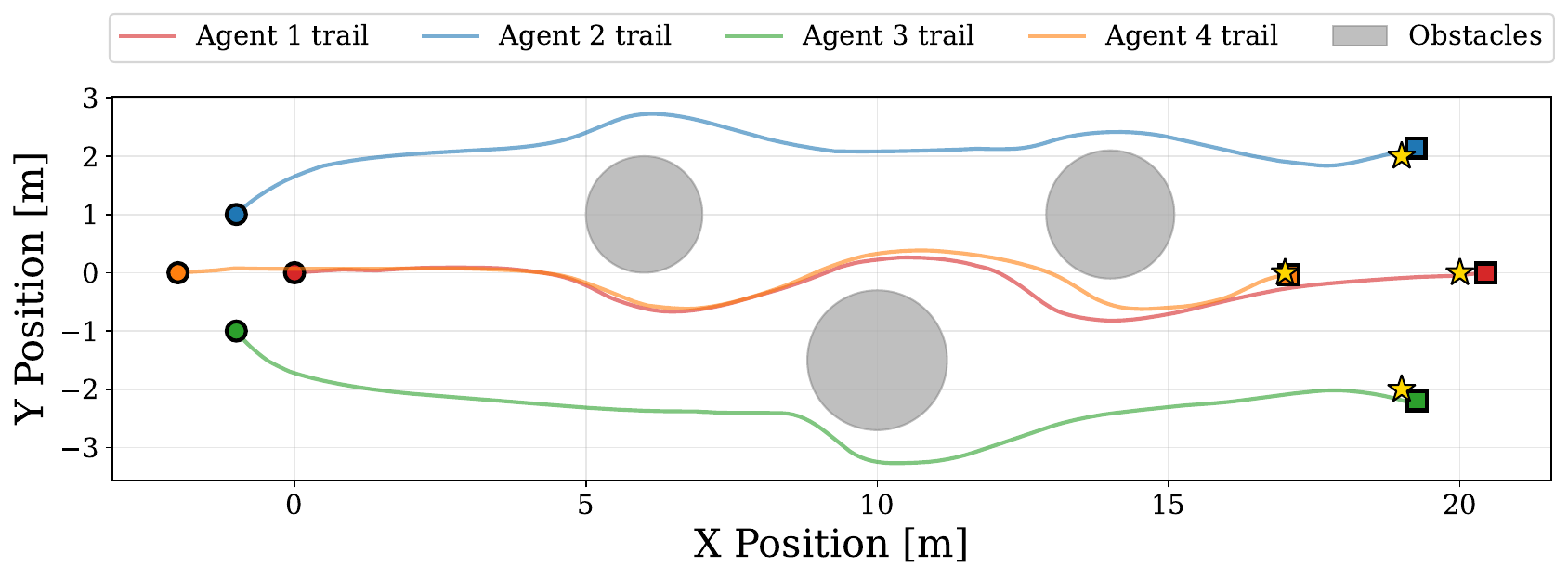}
    \caption{Cooperative}
    \label{fig:traj_coop}
  \end{subfigure}\\
  \begin{subfigure}[b]{\linewidth}
    \centering
    \includegraphics[width=\linewidth]{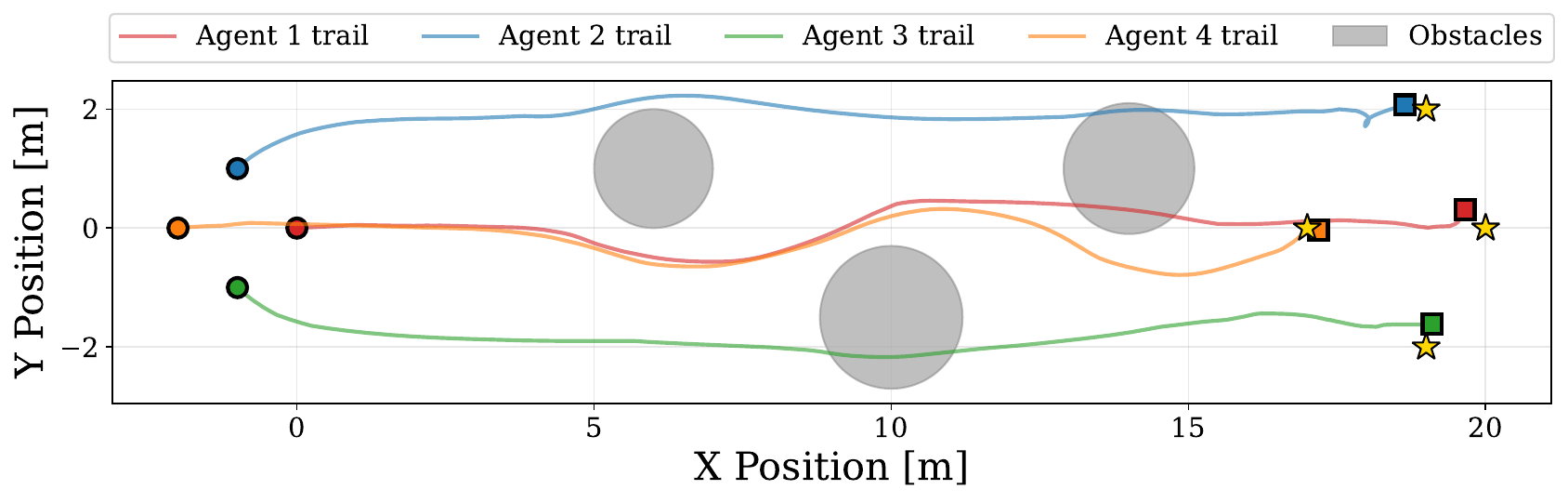}
    \caption{Non-cooperative}
    \label{fig:traj_noncoop}
  \end{subfigure}
  \caption{Agent trajectories under cooperative vs.\ non-cooperative control.}
  \label{fig:traj_compare}
\end{figure}

\begin{figure}[t]
  \centering
  
  \begin{subfigure}[b]{\linewidth}
    \centering
    \includegraphics[width=\linewidth]{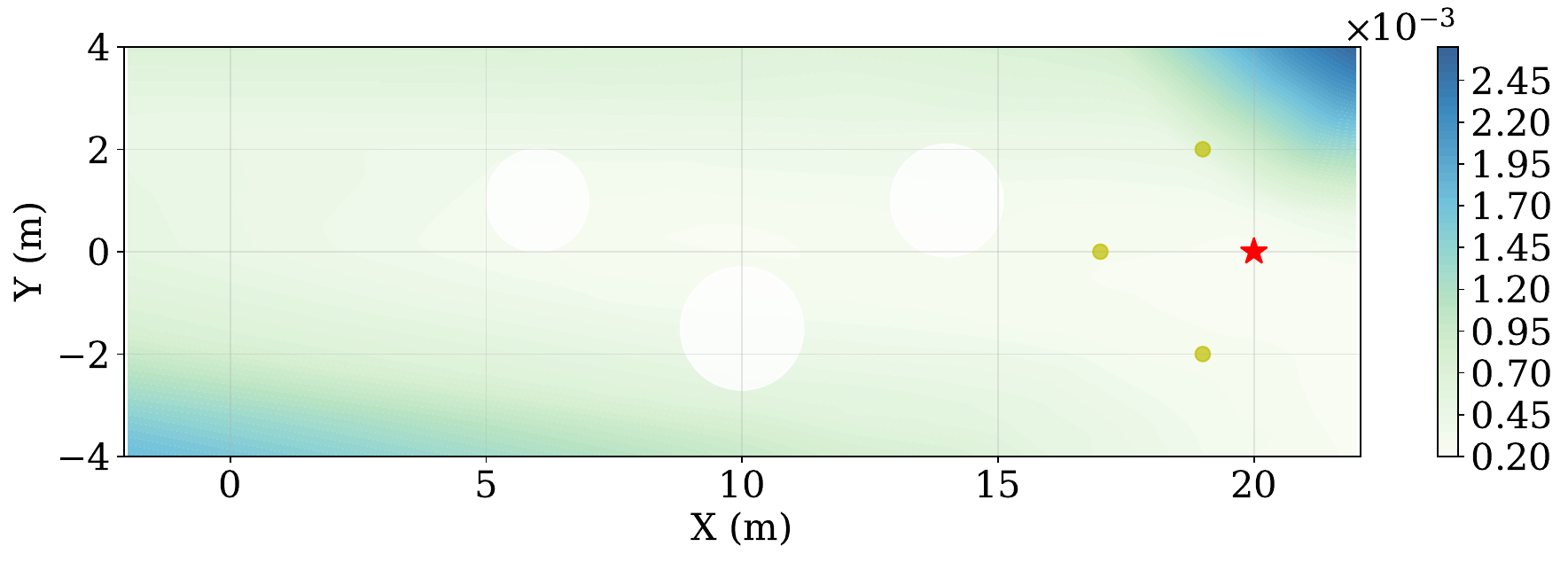}
    \caption{Lyapunov certificate}
    \label{fig:lyap_global}
  \end{subfigure}
  
  \begin{subfigure}[b]{\linewidth}
    \centering
    \includegraphics[width=\linewidth]{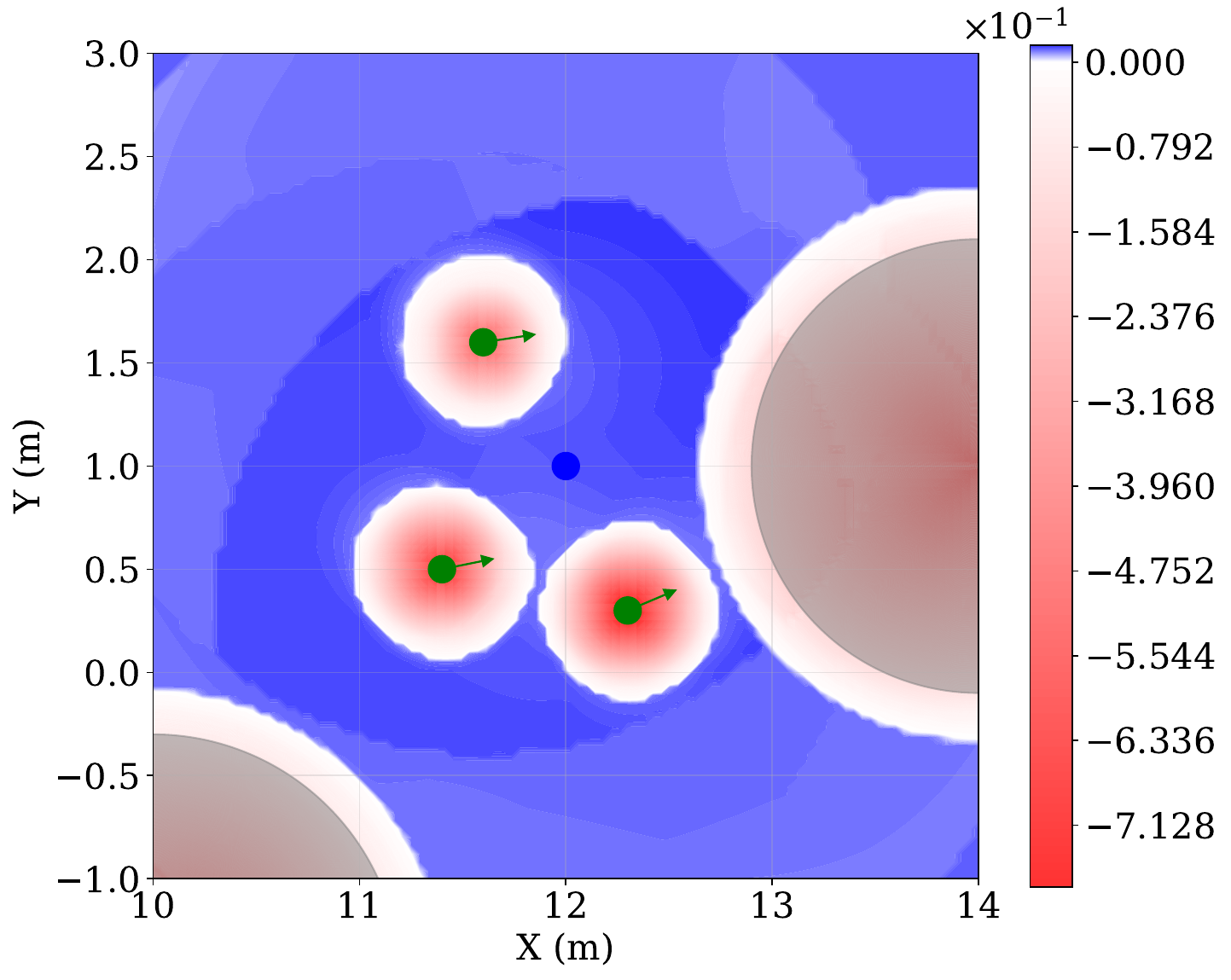}
    \caption{Barrier certificate}
    \label{fig:barrier_local}
  \end{subfigure}
  \caption{Learned certificates for Agent 0.}
  \label{fig:certificates}
\end{figure}

\begin{figure}[h]
  \centering
  \begin{subfigure}[b]{\linewidth}
    \centering
    \includegraphics[width=\linewidth]{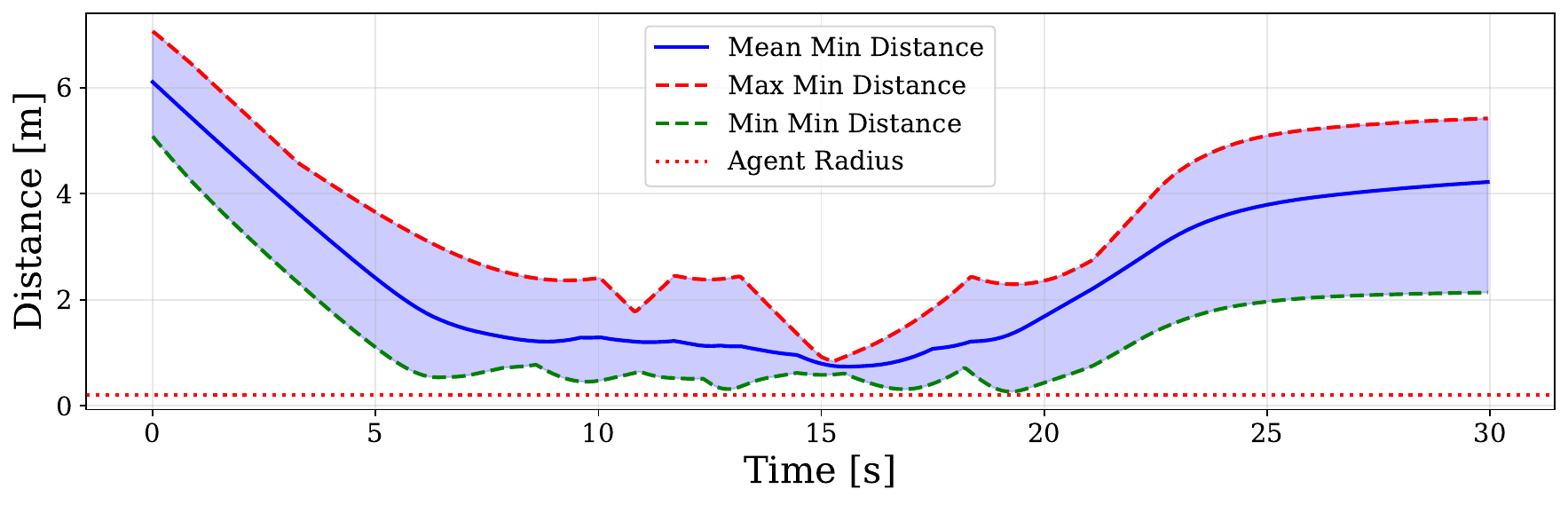}
    \caption{Cooperative}
    \label{fig:obsdist_coop}
  \end{subfigure}\\
  \begin{subfigure}[b]{\linewidth}
    \centering
    \includegraphics[width=\linewidth]{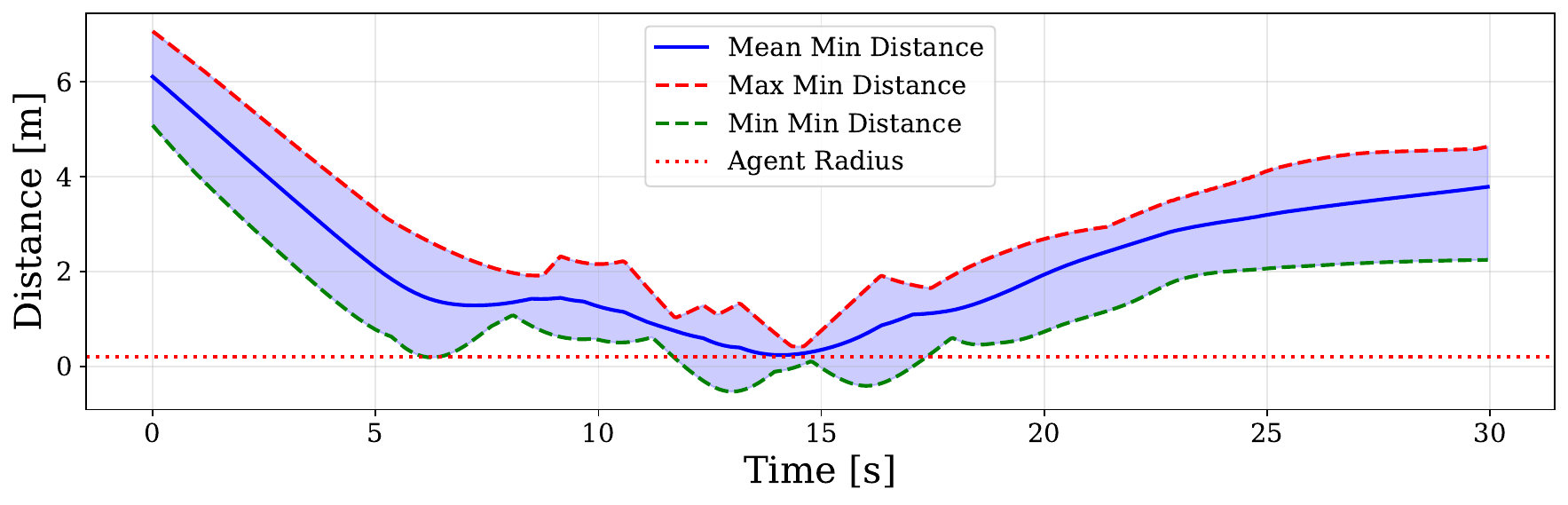}
    \caption{Non-cooperative}
    \label{fig:obsdist_noncoop}
  \end{subfigure}
  \caption{Minimum distance to obstacles over time.}
  \label{fig:obsdist_compare}
\end{figure}

\begin{figure}[h]
  \centering
  \begin{subfigure}[b]{\linewidth}
    \centering
    \includegraphics[width=\linewidth]{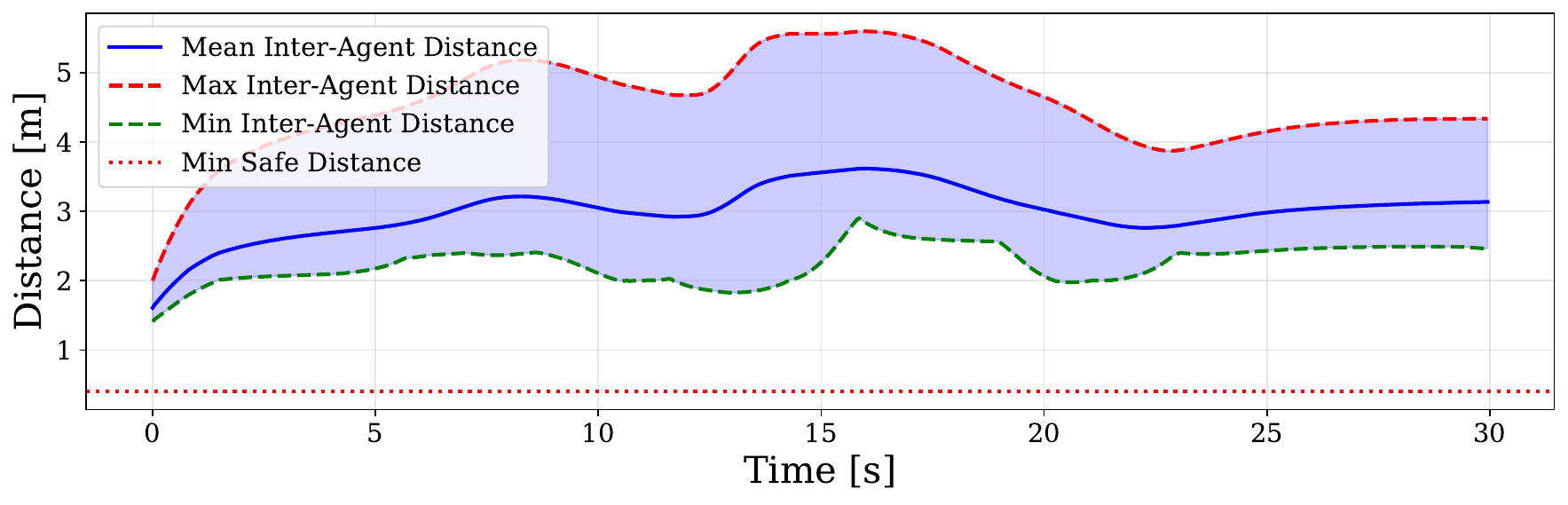}
    \caption{Cooperative}
    \label{fig:iadist_coop}
  \end{subfigure}\\
  \begin{subfigure}[b]{\linewidth}
    \centering
    \includegraphics[width=\linewidth]{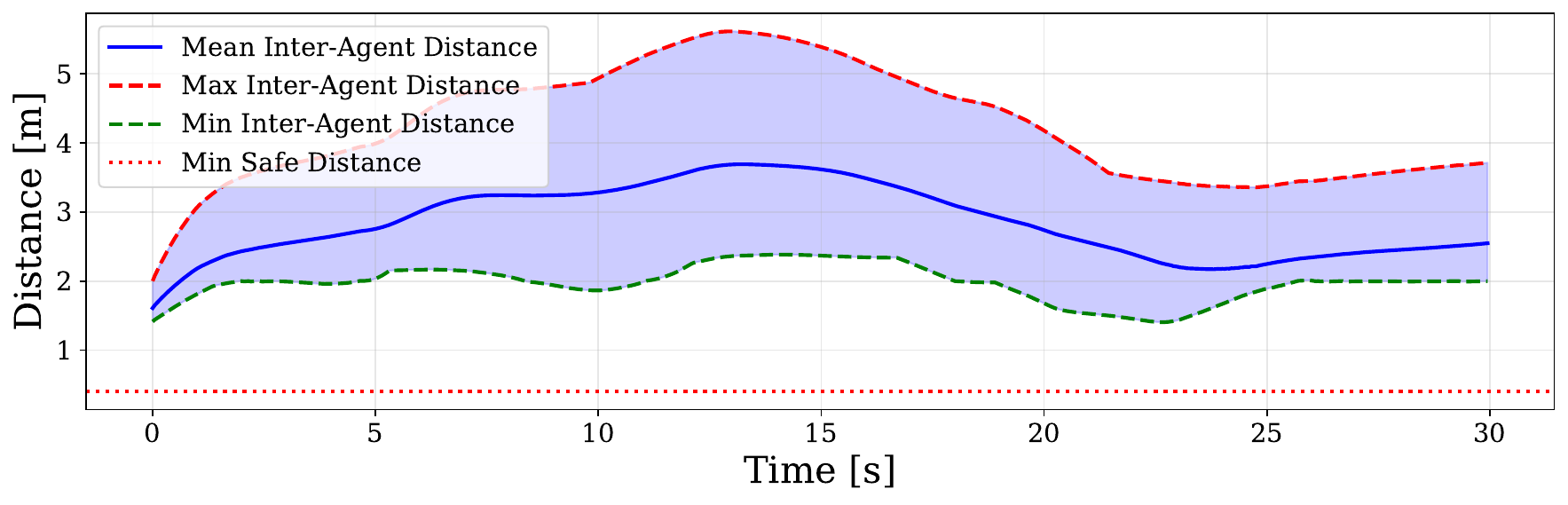}
    \caption{Non-cooperative}
    \label{fig:iadist_noncoop}
  \end{subfigure}
  \caption{Inter-agent distance statistics.}
  \label{fig:iadist_compare}
\end{figure}

To evaluate the effectiveness of the proposed Co-RWA framework, we compare our method against non-cooperative baseline that employ scalar Lyapunov and barrier functions for individual agents as adopted in~\cite{dai2021lyapunov,zhang2025gcbf+} in the aforementioned two scenarios. The baseline treat each agent independently, without modeling inter-agent coupling in the certificates. 

\noindent\textbf{Control Performance for Omnidirectional Robots:}
Figures~\ref{fig:traj_compare}–\ref{fig:iadist_compare} illustrate the control performance of omnidirectional robots under cooperative and non-cooperative settings. As shown in Figures~\ref{fig:traj_compare}, \ref{fig:obsdist_compare}, and~\ref{fig:iadist_compare}, the controller trained with the proposed Co-RWA framework successfully guides agents to their goals while avoiding obstacles and maintaining safe inter-agent distances. In contrast, for the baseline method, it is challenging to obtain valid scalar certificates due to intractability in large-scale interconnected systems. Figure~\ref{fig:certificates} shows the learned Co-RWA certificates for a representative agent, demonstrating well-structured Lyapunov and barrier functions that enforce both liveness and safety through the Co-RWA certificate.

\noindent\textbf{Control Performance for Vehicle Platoons:} Table~\ref{tab:rwa-platoon} summarizes the reach–while–avoid performance of different controllers in the vehicle platoon scenario, including (i) a nominal controller trained with MAPPO (Nominal) and (ii) a non-cooperative controller using neural certificates without coordination (Non-cooperative). 
We report three metrics: (i) \textbf{Tracking Error}, computed as the root mean square error (RMSE) between actual and desired velocity; and (ii) \textbf{Average TTC}~\cite{minderhoud2001extended}, the mean time-to-collision indicating safety margins.

Table~\ref{tab:rwa-platoon} shows that the proposed Co-RWA certificates with the fine-tuned controller achieve a better balance between safety and control performance. It obtains the highest average TTC of 27.88, indicating improved safety margins compared to both the original and non-cooperative controllers. Despite a slightly higher tracking error than the original controller, it significantly outperforms the non-cooperative method, suggesting that Co-RWA certificates help preserve tracking performance while ensuring safety. 

\begin{table}[htp]
\centering
\small
\renewcommand{\arraystretch}{1.15}
\begin{tabular}{lccc}
\toprule
\textbf{Metric} & \textbf{Original} & \textbf{Non-Cooperative} & \textbf{Cooperative} \\ \midrule
Tracking Error & 0.3844 & 0.4354 & 0.4294\\ 
Average TTC & 27.3826 & 24.0206&27.8819\\
\bottomrule
\end{tabular}
\caption{Control performance for vehicle platoons}
\label{tab:rwa-platoon}
\end{table}

\subsection{Verification Performance Analysis}
\label{subsec: verification results}
\begin{table}[h]
\centering
\renewcommand{\arraystretch}{1.15}
\small
\begin{tabular}{@{}llc@{}}
\toprule
\textbf{Method} & \textbf{Platoon Size $\mathcal{N}$} & \textbf{Training Time [s]} \\ \midrule
\multirow{4}{*}{Full R.} 
  & 3   & 2223.3 $\pm$ 313.5 \\
  & 6   & 7207.3 $\pm$ 633.5 \\
  & 30  & Timeout \\
  & 300 & Timeout \\ \midrule
\multirow{4}{*}{RedVer} 
  & 3   & \multirow{4}{*}{\textbf{2223.3 $\pm$ 313.5}} \\
  & 6   &  \\
  & 30  &  \\
  & 300 &  \\
\bottomrule
\end{tabular}
\caption{Scalability study: training and verification time across different platoon sizes}
\label{tab:train_scalability_final}
\end{table}

To evaluate the scalability and efficiency of our framework, we compare two verification strategies: (i) Full Retraining (\textbf{Full R.}), which retrains the controller and neural certificates from scratch for each platoon size $\mathcal{N}$, and (ii) Reduced Verification (\textbf{RedVer}), which trains once on a small platoon ($\mathcal{N}=3$) and reuses the resulting controller and certificates for larger platoons by exploiting structural isomorphisms. We use Marabou~\cite{wu2024marabou} as the verification tool.

These results demonstrate the significant scalability benefits of our RedVer strategy. As shown in Table~\ref{tab:train_scalability_final}, the Full R. strategy exhibits poor scalability: training and verification time increases drastically with $\mathcal{N}$, and fails to terminate for $\mathcal{N} = 30$ and $300$ due to computational limits. In contrast, RedVer achieves constant training and verification cost across all platoon sizes, as it only needs to train and verify a subset of the platoon.

\subsection{Ablation Study}
\label{subsec: Ablation}
The ablation study evaluates three choices of neural network structure (MLP, Deepsets~\cite{zaheer2017deep}, and GNN) under an omnidirectional robot environment. As shown in Table~\ref{tab: ablation}, the GNN-based certificates attain the highest mean speed while also maintaining the largest minimum obstacle and inter-agent distances, indicating that explicitly modeling relational structure among agents and obstacles leads to both more efficient and safer behaviors than the MLP and Deepsets baselines.
\begin{table}[h]
\centering
\small
\renewcommand{\arraystretch}{1.15}
\setlength{\tabcolsep}{2pt} 
\begin{tabular}{lccc}
\toprule
\textbf{Metric} & \textbf{MLP} & \textbf{Deepsets} & \textbf{GNN} \\
\hline
Mean Speed              & $0.677 \pm 0.005$ & $0.682 \pm 0.005$ & $0.687 \pm 0.005$ \\
Min Obstacle Dist.   & $0.111 \pm 0.010$ & $0.087 \pm 0.019$ & $0.272 \pm 0.007$ \\
Min Inter-Agent Dist.& $1.266 \pm 0.057$ & $1.265 \pm 0.052$ & $1.267 \pm 0.056$ \\
\bottomrule
\end{tabular}
\caption{Ablation study of neighborhood encoder architectures. The GNN-based certificates achieve the highest mean speed while maintaining the largest minimum obstacle and inter-agent distances.}
\label{tab: ablation}
\end{table}

\section{Conclusion}
\label{sec:conclusion}
This paper tackles the challenge of ensuring liveness and safety for neural network-based controllers in large-scale interconnected systems. First, we introduce Co-RWA certificates for dynamic neighborhoods in interconnected systems.
Second, we develop a scalable training and verification framework that jointly learns controllers and certificates.
Third, experiments on different scenarios demonstrate that our approach achieves strong control performance with safety and liveness guarantees.
Future work will focus on verifying (i) the proposed certificates under delays and environmental disturbances, (ii) their scalability to high-dimensional systems, and (iii) the associated control policies in real-world robotic hardware implementations.

\bibliographystyle{IEEEtran}
\bibliography{ref.bib}

@ARTICLE{zhou2024enhancing,
  author={Zhou, Jingyuan and Yan, Longhao and Yang, Kaidi},
  journal={IEEE Transactions on Intelligent Vehicles}, 
  title={Enhancing System-Level Safety in Mixed-Autonomy Platoon via Safe Reinforcement Learning}, 
  year={2024},
  volume={},
  number={},
  pages={1-13}}

@article{ZHOU2024104885,
title = {A parameter privacy-preserving strategy for mixed-autonomy platoon control},
journal = {Transportation Research Part C: Emerging Technologies},
volume = {169},
pages = {104885},
year = {2024},
issn = {0968-090X},
doi = {https://doi.org/10.1016/j.trc.2024.104885},
author = {Jingyuan Zhou and Kaidi Yang},
}

@article{ren2023vector,
  title={Vector control Lyapunov and barrier functions for safe stabilization of interconnected systems},
  author={Ren, Wei and Li, Jingjie and Xiong, Junlin and Sun, Xi-Ming},
  journal={SIAM Journal on Control and Optimization},
  volume={61},
  number={5},
  pages={3209--3233},
  year={2023},
  publisher={SIAM}
}

@article{ilchmann2009positivity,
  title={On positivity and stability of linear time-varying Volterra equations},
  author={Ilchmann, Achim and Ngoc, Pham Huu Anh},
  journal={Positivity},
  volume={13},
  number={4},
  pages={671},
  year={2009},
  publisher={Springer Nature BV}
}

@article{hirsch2006monotone,
  title={Monotone dynamical systems},
  author={Hirsch, Morris W and Smith, Hal},
  journal={Handbook of differential equations: ordinary differential equations},
  volume={2},
  pages={239--357},
  year={2006},
  publisher={Elsevier}
}

@article{silva2021string,
  title={String stability in microgrids using frequency controlled inverter chains},
  author={Silva, Guilherme F and Donaire, Alejandro and Seron, Maria M and McFadyen, Aaron and Ford, Jason},
  journal={IEEE Control Systems Letters},
  volume={6},
  pages={1484--1489},
  year={2021},
  publisher={IEEE}
}

@inproceedings{zhang2023compositional,
  title={Compositional neural certificates for networked dynamical systems},
  author={Zhang, Songyuan and Xiu, Yumeng and Qu, Guannan and Fan, Chuchu},
  booktitle={Learning for Dynamics and Control Conference},
  pages={272--285},
  year={2023},
  organization={PMLR}
}

@article{yang2023model,
  title={Model-free safe reinforcement learning through neural barrier certificate},
  author={Yang, Yujie and Jiang, Yuxuan and Liu, Yichen and Chen, Jianyu and Li, Shengbo Eben},
  journal={IEEE Robotics and Automation Letters},
  volume={8},
  number={3},
  pages={1295--1302},
  year={2023},
  publisher={IEEE}
}

@article{zhang2025gcbf+,
  title={Gcbf+: A neural graph control barrier function framework for distributed safe multi-agent control},
  author={Zhang, Songyuan and So, Oswin and Garg, Kunal and Fan, Chuchu},
  journal={IEEE Transactions on Robotics},
  year={2025},
  publisher={IEEE}
}

@article{wang2021beta,
  title={Beta-crown: Efficient bound propagation with per-neuron split constraints for neural network robustness verification},
  author={Wang, Shiqi and Zhang, Huan and Xu, Kaidi and Lin, Xue and Jana, Suman and Hsieh, Cho-Jui and Kolter, J Zico},
  journal={Advances in neural information processing systems},
  volume={34},
  pages={29909--29921},
  year={2021}
}

@inproceedings{lopez2023nnv,
  title={NNV 2.0: the neural network verification tool},
  author={Lopez, Diego Manzanas and Choi, Sung Woo and Tran, Hoang-Dung and Johnson, Taylor T},
  booktitle={International Conference on Computer Aided Verification},
  pages={397--412},
  year={2023},
  organization={Springer}
}

@INPROCEEDINGS{10886052,
  author={Tayal, Manan and Zhang, Hongchao and Jagtap, Pushpak and Clark, Andrew and Kolathaya, Shishir},
  booktitle={2024 IEEE 63rd Conference on Decision and Control (CDC)}, 
  title={Learning a Formally Verified Control Barrier Function in Stochastic Environment}, 
  year={2024},
  volume={},
  number={},
  pages={4098-4104},
  doi={10.1109/CDC56724.2024.10886052}}

@INPROCEEDINGS{10591251,
  author={Wang, Xinyu and Knoedler, Luzia and Mathiesen, Frederik Baymler and Alonso-Mora, Javier},
  booktitle={2024 European Control Conference (ECC)}, 
  title={Simultaneous Synthesis and Verification of Neural Control Barrier Functions Through Branch-and-Bound Verification-in-the-Loop Training}, 
  year={2024},
  volume={},
  number={},
  pages={571-578},
  doi={10.23919/ECC64448.2024.10591251}}

@article{zhang2023exact,
  title={Exact verification of relu neural control barrier functions},
  author={Zhang, Hongchao and Wu, Junlin and Vorobeychik, Yevgeniy and Clark, Andrew},
  journal={Advances in neural information processing systems},
  volume={36},
  pages={5685--5705},
  year={2023}
}

@InProceedings{pmlr-v270-hu25a,
  title = 	 {Verification of Neural Control Barrier Functions with Symbolic Derivative Bounds Propagation},
  author =       {Hu, Hanjiang and Yang, Yujie and Wei, Tianhao and Liu, Changliu},
  booktitle = 	 {Proceedings of The 8th Conference on Robot Learning},
  pages = 	 {1797--1814},
  year = 	 {2024},
  volume = 	 {270},
  publisher =    {PMLR},
}

@inproceedings{
mueller2023certified,
title={Certified Training: Small Boxes are All You Need},
author={Mark Niklas Mueller and Franziska Eckert and Marc Fischer and Martin Vechev},
booktitle={The Eleventh International Conference on Learning Representations },
year={2023},
url={https://openreview.net/forum?id=7oFuxtJtUMH}
}

@inproceedings{ding2022novel,
  title={A novel counterexample-guided inductive synthesis framework for barrier certificate generation},
  author={Ding, Mi and Lin, Kaipeng and Lin, Wang and Ding, Zuohua},
  booktitle={2022 IEEE 33rd International Symposium on Software Reliability Engineering (ISSRE)},
  pages={263--273},
  year={2022},
  organization={IEEE}
}

@inproceedings{verinet,
  author       = {Patrick Henriksen and
                  Alessio R. Lomuscio},
  editor       = {Giuseppe De Giacomo and
                  Alejandro Catal{\'{a}} and
                  Bistra Dilkina and
                  Michela Milano and
                  Sen{\'{e}}n Barro and
                  Alberto Bugar{\'{\i}}n and
                  J{\'{e}}r{\^{o}}me Lang},
  title        = {{Efficient Neural Network Verification via Adaptive Refinement and
                  Adversarial Search}},
  booktitle    = {European Conference on Artificial Intelligence},
  volume       = {325},
  pages        = {2513--2520},
  publisher    = {{IOS} Press},
  year         = {2020}
}

@article{xu2020automatic,
  title={{Automatic Perturbation Analysis for Scalable Certified Robustness and Beyond}},
  author={Xu, Kaidi and Shi, Zhouxing and Zhang, Huan and Wang, Yihan and Chang, Kai-Wei and Huang, Minlie and Kailkhura, Bhavya and Lin, Xue and Hsieh, Cho-Jui},
  journal={Advances in Neural Information Processing Systems},
  volume={33},
  pages={1129--1141},
  year={2020}
}

@inproceedings{nnvTwo,
  author       = {Diego Manzanas Lopez and
                  Sung Woo Choi and
                  Hoang{-}Dung Tran and
                  Taylor T. Johnson},
  editor       = {Constantin Enea and
                  Akash Lal},
  title        = {{{NNV} 2.0: The Neural Network Verification Tool}},
  booktitle    = {International Conference on Computer Aided Verification},
  volume       = {13965},
  pages        = {397--412},
  publisher    = {Springer},
  year         = {2023}
}

@inproceedings{zhou2025synthesis,
  title={Synthesis and Verification of String Stable Control for Interconnected Systems via Neural sISS Certificate},
  author={Zhou, Jingyuan and Wu, Haoze and Yan, Longhao and Yang, Kaidi},
  booktitle={ICLR 2025 Workshop: VerifAI: AI Verification in the Wild},
    year={2025}
}

@article{dai2021lyapunov,
  title={Lyapunov-stable neural-network control},
  author={Dai, Hongkai and Landry, Benoit and Yang, Lujie and Pavone, Marco and Tedrake, Russ},
  journal={Robotics: Science and Systems},
  year={2021}
}

@inproceedings{yanglyapunov,
  title={Lyapunov-stable Neural Control for State and Output Feedback: A Novel Formulation},
  author={Yang, Lujie and Dai, Hongkai and Shi, Zhouxing and Hsieh, Cho-Jui and Tedrake, Russ and Zhang, Huan},
  booktitle={Forty-first International Conference on Machine Learning},
year={2024}
}

@article{mandal2024formally,
  title={Formally Verifying Deep Reinforcement Learning Controllers with Lyapunov Barrier Certificates},
  author={Mandal, Udayan and Amir, Guy and Wu, Haoze and Daukantas, Ieva and Newell, Fletcher Lee and Ravaioli, Umberto J and Meng, Baoluo and Durling, Michael and Ganai, Milan and Shim, Tobey and others},
  journal={arXiv preprint arXiv:2405.14058},
  year={2024}
}

@inproceedings{mandal2024safe,
  title={Safe and Reliable Training of Learning-Based Aerospace Controllers},
  author={Mandal, Udayan and Amir, Guy and Wu, Haoze and Daukantas, Ieva and Newell, Fletcher Lee and Ravaioli, Umberto and Meng, Baoluo and Durling, Michael and Hobbs, Kerianne and Ganai, Milan and others},
  booktitle={2024 AIAA DATC/IEEE 43rd Digital Avionics Systems Conference (DASC)},
  pages={1--10},
  year={2024},
  organization={IEEE}
}

@article{guo2016distributed,
  title={Distributed adaptive integrated-sliding-mode controller synthesis for string stability of vehicle platoons},
  author={Guo, Xianggui and Wang, Jianliang and Liao, Fang and Teo, Rodney Swee Huat},
  journal={IEEE Transactions on Intelligent Transportation Systems},
  volume={17},
  number={9},
  pages={2419--2429},
  year={2016},
  publisher={IEEE}
}

@inproceedings{wu2024marabou,
  title={Marabou 2.0: a versatile formal analyzer of neural networks},
  author={Wu, Haoze and Isac, Omri and Zelji{\'c}, Aleksandar and Tagomori, Teruhiro and Daggitt, Matthew and Kokke, Wen and Refaeli, Idan and Amir, Guy and Julian, Kyle and Bassan, Shahaf and others},
  booktitle={International Conference on Computer Aided Verification},
  pages={249--264},
  year={2024},
  organization={Springer}
}

@article{liu2017distributed,
  title={Distributed model predictive control of spatially interconnected systems using switched cost functions},
  author={Liu, Peng and Ozguner, Umit},
  journal={IEEE transactions on automatic control},
  volume={63},
  number={7},
  pages={2161--2167},
  year={2017},
  publisher={IEEE}
}

@article{liu2023reinforcement,
  title={Reinforcement learning-based decentralized control for networked interconnected systems with communication and control constraints},
  author={Liu, Jinliang and Zhang, Nan and Zha, Lijuan and Xie, Xiangpeng and Tian, Engang},
  journal={IEEE Transactions on Automation Science and Engineering},
  volume={21},
  number={3},
  pages={4674--4685},
  year={2023},
  publisher={IEEE}
}

@article{yang2022dynamic,
  title={Dynamic event-sampled control of interconnected nonlinear systems using reinforcement learning},
  author={Yang, Xiong and Xu, Mengmeng and Wei, Qinglai},
  journal={IEEE Transactions on Neural Networks and Learning Systems},
  volume={35},
  number={1},
  pages={923--937},
  year={2022},
  publisher={IEEE}
}

@article{sturz2020distributed,
  title={Distributed control design for heterogeneous interconnected systems},
  author={St{\"u}rz, Yvonne R and Eichler, Annika and Smith, Roy S},
  journal={IEEE transactions on automatic control},
  volume={66},
  number={11},
  pages={5112--5127},
  year={2020},
  publisher={IEEE}
}

@article{schlaginhaufen2021learning,
  title={Learning stable deep dynamics models for partially observed or delayed dynamical systems},
  author={Schlaginhaufen, Andreas and Wenk, Philippe and Krause, Andreas and Dorfler, Florian},
  journal={Advances in Neural Information Processing Systems},
  volume={34},
  pages={11870--11882},
  year={2021}
}

@inproceedings{nadali2024neural,
  title={Neural closure certificates},
  author={Nadali, Alireza and Murali, Vishnu and Trivedi, Ashutosh and Zamani, Majid},
  booktitle={Proceedings of the AAAI Conference on Artificial Intelligence},
  volume={38},
  number={19},
  pages={21446--21453},
  year={2024}
}

@inproceedings{neustroev2025neural,
  title={Neural continuous-time supermartingale certificates},
  author={Neustroev, Grigory and Giacobbe, Mirco and Lukina, Anna},
  booktitle={Proceedings of the AAAI Conference on Artificial Intelligence},
  volume={39},
  number={26},
  pages={27538--27546},
  year={2025}
}

@inproceedings{yu2025neural,
  title={Neural Control and Certificate Repair via Runtime Monitoring},
  author={Yu, Emily and Zikelic, Djordje and Henzinger, Thomas A},
  booktitle={Proceedings of the AAAI Conference on Artificial Intelligence},
  volume={39},
  number={25},
  pages={26409--26417},
  year={2025}
}

@inproceedings{zhao2024neural,
  title={Neural Barrier Certificates Synthesis of NN-Controlled Continuous Systems via Counterexample-Guided Learning},
  author={Zhao, Hanrui and Qi, Niuniu and Ren, Mengxin and Zeng, Xia and Zeng, Zhenbing and Yang, Zhengfeng},
  booktitle={Proceedings of the 61st ACM/IEEE Design Automation Conference},
  pages={1--6},
  year={2024}
}

@inproceedings{qinlearning,
  title={Learning Safe Multi-agent Control with Decentralized Neural Barrier Certificates},
year = {2021},
  author={Qin, Zengyi and Zhang, Kaiqing and Chen, Yuxiao and Chen, Jingkai and Fan, Chuchu},
  booktitle={International Conference on Learning Representations}
}

@article{minderhoud2001extended,
  title={Extended time-to-collision measures for road traffic safety assessment},
  author={Minderhoud, Michiel M and Bovy, Piet HL},
  journal={Accident Analysis \& Prevention},
  volume={33},
  number={1},
  pages={89--97},
  year={2001},
  publisher={Elsevier}
}

@article{hu2022processing,
  title={Processing, assessing, and enhancing the Waymo autonomous vehicle open dataset for driving behavior research},
  author={Hu, Xiangwang and Zheng, Zuduo and Chen, Danjue and Zhang, Xi and Sun, Jian},
  journal={Transportation Research Part C: Emerging Technologies},
  volume={134},
  pages={103490},
  year={2022},
  publisher={Elsevier}
}

@inproceedings{clark2021verification,
  title={Verification and synthesis of control barrier functions},
  author={Clark, Andrew},
  booktitle={2021 60th IEEE Conference on Decision and Control (CDC)},
  pages={6105--6112},
  year={2021},
  organization={Ieee}
}

@article{duan2021graph,
  title={Graph-theoretic stability conditions for Metzler matrices and monotone systems},
  author={Duan, Xiaoming and Jafarpour, Saber and Bullo, Francesco},
  journal={SIAM Journal on Control and Optimization},
  volume={59},
  number={5},
  pages={3447--3471},
  year={2021},
  publisher={SIAM}
}

@book{khalil2002nonlinear,
  title={Nonlinear systems},
  author={Khalil, Hassan K and Grizzle, Jessy W},
  volume={3},
  year={2002},
  publisher={Prentice hall Upper Saddle River, NJ}
}

@inproceedings{tan2021deductive,
  title={Deductive stability proofs for ordinary differential equations},
  author={Tan, Yong Kiam and Platzer, Andr{\'e}},
  booktitle={International Conference on Tools and Algorithms for the Construction and Analysis of Systems},
  pages={181--199},
  year={2021},
  organization={Springer}
}

@book{allen2008mathematical,
  title={Mathematical epidemiology},
  author={Allen, Linda JS and Brauer, Fred and Van den Driessche, Pauline and Wu, Jianhong},
  volume={1945},
  year={2008},
  publisher={Springer}
}

@article{zaheer2017deep,
  title={Deep sets},
  author={Zaheer, Manzil and Kottur, Satwik and Ravanbakhsh, Siamak and Poczos, Barnabas and Salakhutdinov, Russ R and Smola, Alexander J},
  journal={Advances in neural information processing systems},
  volume={30},
  year={2017}
}

@article{xu2024eclipse,
  title={Eclipse: Efficient compositional lipschitz constant estimation for deep neural networks},
  author={Xu, Yuezhu and Sivaranjani, S},
  journal={Advances in Neural Information Processing Systems},
  volume={37},
  pages={10414--10441},
  year={2024}
}

@article{sun2019adaptive,
  title={Adaptive decentralized neural network tracking control for uncertain interconnected nonlinear systems with input quantization and time delay},
  author={Sun, Haibin and Hou, Linlin and Zong, Guangdeng and Yu, Xinghuo},
  journal={IEEE transactions on neural networks and learning systems},
  volume={31},
  number={4},
  pages={1401--1409},
  year={2019},
  publisher={IEEE}
}

@article{zhang2022adaptive,
  title={Adaptive decentralized control for interconnected time-delay uncertain nonlinear systems with different unknown control directions and deferred full-state constraints},
  author={Zhang, Liuliu and Zhu, Lingchen and Hua, Changchun and Qian, Cheng},
  journal={IEEE Transactions on Neural Networks and Learning Systems},
  volume={34},
  number={12},
  pages={10789--10801},
  year={2022},
  publisher={IEEE}
}

@article{gu2021safety,
  title={Safety-critical containment maneuvering of underactuated autonomous surface vehicles based on neurodynamic optimization with control barrier functions},
  author={Gu, Nan and Wang, Dan and Peng, Zhouhua and Wang, Jun},
  journal={IEEE Transactions on Neural Networks and Learning Systems},
  volume={34},
  number={6},
  pages={2882--2895},
  year={2021},
  publisher={IEEE}
}

@article{zhang2022barrier,
  title={Barrier Lyapunov function-based safe reinforcement learning for autonomous vehicles with optimized backstepping},
  author={Zhang, Yuxiang and Liang, Xiaoling and Li, Dongyu and Ge, Shuzhi Sam and Gao, Bingzhao and Chen, Hong and Lee, Tong Heng},
  journal={IEEE Transactions on Neural Networks and Learning Systems},
  volume={35},
  number={2},
  pages={2066--2080},
  year={2022},
  publisher={IEEE}
}

@article{fuentes2020adaptive,
  title={Adaptive tracking control of state constraint systems based on differential neural networks: A barrier Lyapunov function approach},
  author={Fuentes-Aguilar, Rita Q and Chairez, Isaac},
  journal={IEEE transactions on neural networks and learning systems},
  volume={31},
  number={12},
  pages={5390--5401},
  year={2020},
  publisher={IEEE}
}

@article{wei2022stability,
  title={Stability of delayed reaction--diffusion neural-network models with hybrid impulses via vector Lyapunov function},
  author={Wei, Tengda and Li, Xiaodi and Cao, Jinde},
  journal={IEEE Transactions on Neural Networks and Learning Systems},
  volume={34},
  number={10},
  pages={7467--7478},
  year={2022},
  publisher={IEEE}
}

@article{cai2023fixed,
  title={Fixed-time control and estimation of discontinuous fuzzy neural networks: Novel Lyapunov method of fixed-time stability},
  author={Cai, Zuowei and Huang, Lihong and Wang, Zengyun},
  journal={IEEE Transactions on Neural Networks and Learning Systems},
  year={2023},
  publisher={IEEE}
}
\appendix

\subsection{Proof for Theorem~\ref{thm:dl-vclf-stab}}
\label{app: theorem 1}
\begin{proof}
We prove Theorem~\ref{thm:dl-vclf-stab} by constructing a scalar Lyapunov candidate written as follows:
\begin{equation}
V_p(\bm{x}) := p^\top V(\bm{x})
\end{equation}
where since $\Lambda$ is Metzler and Hurwitz, $p \in \mathbb{R}_{>0}^q$ is chosen to be a  strictly positive vector such that 
\begin{equation}
p^\top \Lambda = -c^\top
\end{equation}
for some vector $c \in \mathbb{R}_{>0}^q$ (see Theorem 5.2.1 in \cite{allen2008mathematical}). 

We next use $V_p$ to show the exponential stability. By the DL-VCLF condition and the definition of $W_i(\bm{x})$,
\begin{equation}
\dot V_i(x_i) = L_{f_i}V_i\bigl(x_i\bigr)
+ L_{g_i}V_i(x_i)\,u_i(\bar{x}_i)
\le W_i^\top(\bm{x})\,V(\bm{x})
\end{equation}
and since $W_i(\bm{x}) \le \lambda_i$ (component-wise), stacking across all agents yields
\begin{equation}
\begin{aligned}
\dot V(\bm{x})
&= \begin{bmatrix} \dot V_1(x_1) \\ \vdots \\ \dot V_q(x_q) \end{bmatrix}
\le
\begin{bmatrix} W_1^\top(\bm{x}) V(\bm{x}) \\ \vdots \\ W_q^\top(\bm{x}) V(\bm{x}) \end{bmatrix}
\le
\begin{bmatrix} \lambda_1^\top V(\bm{x}) \\ \vdots \\ \lambda_q^\top V(\bm{x}) \end{bmatrix}
\\&= \Lambda V(\bm{x}).
\end{aligned}
\label{eq:Vdot_bound}
\end{equation}
Multiplying both sides by the strictly positive vector $p^\top$, we get
\begin{equation}
\begin{aligned}
\dot V_p(\bm{x})
&= p^\top \dot V(\bm{x})
\;\le\; p^\top \Lambda V(\bm{x})
= -c^\top V(\bm{x}) \\
&=-\sum_{i}c_iV_i(x_i)=-\sum_i\frac{c_i}{p_i}p_iV_i(x_i)\\
&\le\; -\min_i\frac{c_i}{p_i} \cdot p^\top V(\bm{x})
= -c_{\min} V_p(\bm{x}).
\end{aligned}
\end{equation}
where $c_{\min} := \min_i \frac{c_i}{p_i} > 0$.

This implies the exponential decay:
\begin{equation}
V_p(\bm{x}(t)) \le V_p(\bm{x}(0))\,e^{-c_{\min} t}, \quad \forall t \ge 0.
\end{equation}

Since each $V_i(x_i)$ is positive definite and radially unbounded, their weighted sum \( V_p(\bm{x}) \) satisfies
\begin{equation}
\alpha_1(\|\bm{x}\|) \le V_p(\bm{x}) \le \alpha_2(\|\bm{x}\|)
\end{equation}
for some class-$\mathcal{K}_\infty$ functions $\alpha_1,\alpha_2$, implying \( \|\bm{x}(t)\|\to 0 \) exponentially.

Thus, the equilibrium is globally exponentially stable.
\end{proof}

\subsection{Proof for Theorem~\ref{thm:dl-vblf-safe}}
\label{app: theorem 2}
\begin{proof}
Let the Lipschitz feedbacks \(u_i(\bar x_i)\) be chosen such that, for every
\(i\in\mathcal N\),
\begin{equation}\label{eq:DLVCBF-cond}
  L_{f_i}h_i(\bar x_i)\;+\;L_{g_i}h_i(\bar x_i)\,u_i(\bar x_i)
  \;\ge\;
  \Gamma_i(\bm{x})^{\!\top}\,h(\bm{x}),
\end{equation}
where each \(\Gamma_i(\bm{x})=\mu_i\circ A_i(\bm{x})\) and the stacked matrix
\(\Gamma(\bm{x}):=(\Gamma_1(\bm{x}),\dots,\Gamma_q(\bm{x}))\) is Metzler.

Then we define the comparison system
\begin{equation}\label{eq:comp}
  \bm{\dot z}(t)\;=\;\Gamma\!\bigl(\bm{z}(t)\bigr)\,\bm{z}(t).
\end{equation}
where we set $\bm{z}(0)\;:=\;h(\bm{x}(0))\succeq0$. We will next show the forward invariance of the comparison system and further of the safe set $\mathcal{C}$. 

As $\Gamma(\bm{x}(t))$ is Metzler for every $t$, system~\eqref{eq:comp}
is a \emph{time–varying positive system}.
Indeed, it is a particular case of the Volterra equation
\begin{equation}
   \bm{\dot z}(t)=A(t)\bm{z}(t)+\int_{t_0}^{t}B(t,s)\bm{z}(s)\,ds
\end{equation}
with kernel \(B\equiv0\) and \(A(t)=\Gamma(\bm{z}(t))\).  
By Theorem 2.4 of \cite{ilchmann2009positivity}, any solution of~\eqref{eq:comp} that starts in the positive orthant
remains in the positive orthant:
\begin{equation}\label{eq:z-nonneg}
  \bm{z}(0)\succeq0\quad\Longrightarrow\quad \bm{z}(t)\succeq0,\;\;\forall\,t\ge0.
\end{equation}

From~\eqref{eq:DLVCBF-cond}, we have the differential inequality
\[
  \dot h(\bm{x})\;\succeq\;\Gamma\!\bigl(\bm{x}(t)\bigr)\,h(\bm{x}).
\]
Because both the true system for \(h\) and the comparison
system~\eqref{eq:comp} share the same Metzler right–hand side,
the vector comparison principle for cooperative systems
~\cite{hirsch2006monotone}
implies
\begin{equation}
  h(\bm{x}(0))=\bm{z}(0)\quad\Longrightarrow\quad
  h\bigl(\bm{x}(t)\bigr)\;\succeq\;\bm{z}(t),\;\;\forall\,t\ge0.
\end{equation}

Combining this with~\eqref{eq:z-nonneg} yields
\begin{equation}
  h\bigl(\bm{x}(t)\bigr)\;\succeq\;0,\qquad\forall\,t\ge0.
\end{equation}

Then, if the initial state satisfies $\bm{x}(0)\in\mathcal C$
(i.e., $h_i(\bar x_i(0))\ge0$ for all $i$),
then the inequality above shows $h_i(\bar x_i(t))\ge0$
for every $t\ge0$, hence $x(t)\in\mathcal C$ for all time.
Therefore the safe set $\mathcal C$ is forward invariant under the
control law $\bm{u}=(u_1,\dots,u_q)$.
\end{proof}

\subsection{Proof for Theorem~\ref{thm:disc-CoRWA}}
\begin{proof}
We first prove Theorem~\ref{thm:disc-CoRWA} by quantifying (i) the discretization error and (ii) the approximation error, and then incorporating these errors into~\eqref{eq:VCLF_condition} and~\eqref{eq: CBF}.

For discretization error, we fix $i\in\mathcal N$ and $k\in\mathbb N$. For $s\in[kT,(k+1)T]$, by definition
\begin{align}
\dot x_i(s) = f_i(\bar x_i(s))+g_i(\bar x_i(s))u_i(s),
\end{align}
and Assumption~\ref{asmp: Lipschitz} implies
\begin{align}
\|\dot x_i(s)\|
\;\le\; \sup_{\bar x_i,u_i}\bigl\|f_i(\bar x_i)+g_i(\bar x_i)u_i\bigr\|
= M_{x_i} < \infty .
\end{align}
Hence
\begin{align}
\|x_i(s)-x_i(kT)\|
\le \int_{kT}^{s}\|\dot x_i(\tau)\|\,d\tau
\le M_{x_i}(s-kT).
\label{eq:state-variation}
\end{align}

With the Euler one–step approximation and its local truncation error
\begin{align}
\hat x_i((k+1)T)
&= x_i(kT) + T\Bigl(f_i(\bar x_i(kT))
+\notag\\
&g_i(\bar x_i(kT))u_i(kT)\Bigr),\\
e_i(k) &= \hat x_i((k+1)T) - x_i((k+1)T).
\end{align}
Using the integral form of the solution,
\begin{align}
&e_i(k)
= \int_{kT}^{(k+1)T}
\Bigl[ f_i(\bar x_i(kT))+g_i(\bar x_i(kT))u_i(kT)\notag\\
&\hspace{7em}-f_i(\bar x_i(s))-g_i(\bar x_i(s))u_i(s)
\Bigr]\,ds.
\end{align}
By Lipschitz continuity of $f_i$ and $g_i$ in $\bar x_i$ and boundedness of $u_i$, there exists a constant $L_{x_i}>0$ such that
\begin{align}
&\bigl\|f_i(\bar x_i(s))+g_i(\bar x_i(s))u_i(s)
      -f_i(\bar x_i(kT))\notag\\
&-g_i(\bar x_i(kT))u_i(kT)\bigr\|
\le L_{x_i}\|\bar x_i(s)-\bar x_i(kT)\|.
\end{align}
Similar to~\eqref{eq:state-variation} and by the definition of $\bar x_i$, we have 
\begin{align}
\|\bar x_i(s)-\bar x_i(kT)\|
&\le \bar M_{x_i}(s-kT),
\end{align}
where
\begin{align}
\bar M_{x_i}
:= \Biggl(\sum_{j\in{i}\cup\mathcal E_i} M_{x_j}^2\Biggr)^{1/2}.
\end{align}
and therefore
\begin{align}
&\|e_i(k)\|
\le \int_{kT}^{(k+1)T} L_{x_i}\|\bar x_i(s)-\bar x_i(kT)\|\,ds\notag\\
&\le L_{x_i}\bar{M}_{x_i}\int_{kT}^{(k+1)T}(s-kT)\,ds
= \tfrac{1}{2}L_{x_i}\bar{M}_{x_i}T^2. \label{eq:euler-error}
\end{align}

Next, we bound the errors due to discretization \eqref{eq: discrete_system} as 
\begin{align}
\Delta_i^{\mathrm{disc}}(k)
:= \frac{V_i(\hat x_i((k+1)T))-V_i(x_i(kT))}{T}
\end{align}
and the continuous-time Lie derivative
\begin{align}
\dot V_i(x_i(kT))
:= \nabla V_i(x_i(kT))^\top\dot x_i(kT).
\end{align}

First, by adding and subtracting $V_i(x_i((k+1)T))$ and using the Lipschitz
constant $L_{V_i}$ of $V_i$,
\begin{align}
&\left|\frac{V_i(\hat x_i((k+1)T))-V_i(x_i((k+1)T))}{T}\right|
\le \frac{L_{V_i}}{T}\,\|e_i(k)\|\notag\\
&\le \tfrac{1}{2}L_{V_i}L_{x_i}\bar{M}_{x_i}T. \label{eq:V-part}
\end{align}
Second, using the fundamental theorem of calculus,
\begin{align}
&\frac{V_i(x_i((k+1)T))-V_i(x_i(kT))}{T}
= \frac{1}{T}\int_{kT}^{(k+1)T}\dot V_i(x_i(s))\,ds\notag\\
&=: \dot{\widetilde V}_i(x_i(kT)).
\end{align}
Since $\dot V_i$ is Lipschitz with constant $L_{\dot V_i}$, we have
\begin{align}
|\dot V_i(x_i(s))-\dot V_i(x_i(kT))|
&\le L_{\dot V_i}\|x_i(s)-x_i(kT)\|\notag\\
&\le L_{\dot V_i}\bar{M}_{x_i}(s-kT),
\end{align}
where we used~\eqref{eq:state-variation}. Consequently,
\begin{align}
\bigl|\dot{\widetilde V}_i(x_i(kT))-\dot V_i(x_i(kT))\bigr|
&\le \frac{1}{T}\int_{kT}^{(k+1)T}
      L_{\dot V_i}\bar{M}_{x_i}(s-kT)\,ds\notag\\
&= \tfrac{1}{2}T L_{\dot V_i}\bar{M}_{x_i}. \label{eq:dV-part}
\end{align}

Combining \eqref{eq:V-part} and \eqref{eq:dV-part} yields
\begin{align}
\bigl|\Delta_i^{\mathrm{disc}}(k)-\dot V_i(x_i(kT))\bigr|
&\le \tfrac{1}{2}T\bigl(L_{V_i}L_{x_i} + L_{\dot V_i}\bigr)\bar{M}_{x_i}.
\end{align}
Then we get the error bound for discretization as:
\begin{align}
e^{V,\mathrm{disc}}_{i,\sup}:=\tfrac{1}{2}T\bigl(L_{V_i}L_{x_i} + L_{\dot V_i}\bigr)\bar{M}_{x_i}. \label{eq:disc-only-bound}
\end{align}

Then we calculate the approximation error. For simplicity, let $\tilde F_i(\bar x_i,u_i)
:= \tilde f_i(\bar x_i) + \tilde g_i(\bar x_i)u_i(\bar{x}_i)$ denote the surrogate closed-loop dynamics, and define the Euler one-step
approximation generated by $\tilde F_i$ as
\begin{align}
\tilde x_i((k+1)T)
:= x_i(kT) + T\,\tilde F_i(\bar x_i(kT),u_i(kT)).
\end{align}
By Assumption~\ref{asmp: uniform sample},
\begin{align}
\bigl\|F_i(\bar x_i(kT),u_i(kT)) - \tilde F_i(\bar x_i(kT),u_i(kT))\bigr\|
\le \hat{\epsilon}_i,
\end{align}
where $F_i(\bar x_i,u_i) := f_i(\bar x_i)+g_i(\bar x_i)u_i$, and hence
\begin{align}
&\bigl\|\tilde x_i((k+1)T) - \hat x_i((k+1)T)\bigr\|
\notag\\&= T
\bigl\|F_i(\bar x_i(kT),u_i(kT)) - \tilde F_i(\bar x_i(kT),u_i(kT))\bigr\|
\le T\hat{\epsilon}_i.
\end{align}
Using the Lipschitz constant $L_{V_i}$ of $V_i$, we obtain
\begin{align}
\left|
\frac{V_i(\tilde x_i((k+1)T)) - V_i(\hat x_i((k+1)T))}{T}
\right|
\le L_{V_i}\hat{\epsilon}_i
=: e^{V,\mathrm{approx}}_{i,\sup}. \label{eq:approx-V-part}
\end{align}
Define the discrete-time term based on the surrogate dynamics as
\begin{align}
\Delta_i^{\mathrm{NN}}(k)
:= \frac{V_i(\tilde x_i((k+1)T)) - V_i(x_i(kT))}{T}.
\end{align}
By the triangle inequality and \eqref{eq:disc-only-bound}–\eqref{eq:approx-V-part},
\begin{align}
&\bigl|\Delta_i^{\mathrm{NN}}(k) - \dot V_i(x_i(kT))\bigr|\notag\\
&\le \bigl|\Delta_i^{\mathrm{NN}}(k) - \Delta_i^{\mathrm{disc}}(k)\bigr|
    + \bigl|\Delta_i^{\mathrm{disc}}(k) - \dot V_i(x_i(kT))\bigr|\notag\\
&\le e^{V,\mathrm{approx}}_{i,\sup} + e^{V,\mathrm{disc}}_{i,\sup}\notag\\
&=: e^{V}_{i,\sup}. \label{eq:total-V-error}
\end{align}

Now assume the discrete-time inequality in Theorem~\ref{thm:disc-CoRWA},
\begin{align}
\Delta_i^{\mathrm{NN}}(k)
\le \lambda_i^\top V(x(kT)) - e^{V}_{i,\sup},
\end{align}
holds for all $k$ and $i$. Using the bound~\eqref{eq:total-V-error},
\begin{align}
\dot V_i(x_i(kT))
\le \Delta_i^{\mathrm{NN}}(k) + e^{V}_{i,\sup}
\le \lambda_i^\top V(x(kT)),
\end{align}
which is exactly the continuous-time Co-RWA matrix inequality for $V_i$.

The argument for the barrier functions $h_i$ is completely analogous:
replacing $V_i$ by $h_i$, $L_{\dot V_i}$ by $L_{\dot h_i}$, and $L_{V_i}$
by $L_{h_i}$ gives
\begin{align}
&\left|
\frac{h_i(\tilde{\bar x}_i((k+1)T))-h_i(\bar x_i(kT))}{T}
- \dot h_i(\bar x_i(kT))
\right|
\le e^{h}_{i,\sup},
\end{align}
with $e^{h}_{i,\sup}$ as in Theorem~\ref{thm:disc-CoRWA}, so the discrete
inequality
\begin{align}
\frac{h_i(\tilde{\bar x}_i((k+1)T))-h_i(\bar x_i(kT))}{T}
\ge \mu_i^\top h(x(kT)) + e^{h}_{i,\sup}
\end{align}
implies
$\dot h_i(\bar x_i(kT)) \ge \mu_i^\top h(x(kT))$
for all $i$.

Thus, all continuous-time Co-RWA inequalities in
Def.~\ref{def:CoRWA} hold, and
$\{V_i,h_i\}_{i\in\mathcal N}$ form a neural Co-RWA certificate.
\end{proof}

\subsection{Proof for Theorem~\ref{thm:RWA-Equivalence}}
\label{app: theorem 3}
\begin{proof}
Since $\widetilde{\mathcal{I}}$ is substructure-isomorphic to $\mathcal{I}$, there exists an injective map $\tau: \widetilde{\mathcal{N}} \to \mathcal{N}$ such that, for all $x \in \mathbb{R}^{nq}$ and $j \in \widetilde{\mathcal{N}}$,
\begin{equation}
\tilde{f}_j = f_{\tau(j)}, \quad \tilde{\mathcal{N}}_j(\bm{x}) = \mathcal{N}_{\tau(j)}(\bm{x}),
\end{equation}
and we assume that the local state $\tilde{x}_j$ of agent $j$ in $\widetilde{\mathcal{I}}$ matches that of agent $\tau(j)$ in $\mathcal{I}$ under the same input $x$ (i.e., $\tilde{x}_j = x_{\tau(j)}$).

Let $\tilde{\pi}_j := \pi_{\tau(j)}$, $\tilde{V}_j := V_{\tau(j)}$, and $\tilde{h}_j := h_{\tau(j)}$ be the copied controller and certificates.

\paragraph{DL-VCLF Condition.}
By assumption, the DL-VCLF condition holds in system $\mathcal{I}$:
\begin{equation}
L_{f_{\tau(j)}} V_{\tau(j)} + L_{g_{\tau(j)}} V_{\tau(j)} \pi_{\tau(j)} \leq W_{\tau(j)}^\top V,
\end{equation}
where $W_{\tau(j)}(x) = \lambda_{\tau(j)} \circ A_{\tau(j)}(x)$ is defined using a Metzler matrix $\Lambda = (\lambda_i^\top)_{i \in \mathcal{N}}$.

By construction, since $\tilde{f}_j = f_{\tau(j)}$, $\tilde{\pi}_j = \pi_{\tau(j)}$, and $\tilde{V}_j = V_{\tau(j)}$, the same inequality holds in system $\widetilde{\mathcal{I}}$, with the interaction weights restricted to the subgraph $\widetilde{\mathcal{N}}$.

We define $\tilde{W}_j$ as the truncated vector obtained by selecting entries of $W_{\tau(j)}$ corresponding to agents in $\widetilde{\mathcal{N}}$, i.e.,
\begin{equation}
\tilde{W}_j := \left( w_{\tau(j), \tau(m)} \right)_{m \in \widetilde{\mathcal{N}}}.
\end{equation}
Then the transferred inequality becomes
\begin{equation}
L_{\tilde{f}_j} \tilde{V}_j + L_{\tilde{g}_j} \tilde{V}_j \tilde{\pi}_j \leq \tilde{W}_j^\top \tilde{V},
\end{equation}
where $\tilde{V} = (\tilde{V}_j)_{j \in \widetilde{\mathcal{N}}}$.

Since $\Lambda$ is Metzler and Hurwitz, and $\widetilde{\Lambda}$ is the principal submatrix corresponding to a substructure-isomorphic subsystem with preserved neighborhood topology under $\tau$, the , $\widetilde{\Lambda}$ remains Metzler and Hurwitz. This follows from the closure of Hurwitz stability in Metzler matrices under structurally consistent subgraphs~\cite{duan2021graph}[Lemma~4.1].

Therefore, the DL-VCLF condition is satisfied in $\widetilde{\mathcal{I}}$, ensuring exponential stability.

\paragraph{DL-VCBF Condition.}
Similarly, the DL-VCBF condition in $\mathcal{I}$ satisfies
\begin{equation}
L_{f_{\tau(j)}} h_{\tau(j)} + L_{g_{\tau(j)}} h_{\tau(j)} \pi_{\tau(j)} \geq \Gamma_{\tau(j)}^\top h,
\end{equation}
where $\Gamma_{\tau(j)}(x) = \mu_{\tau(j)} \circ A_{\tau(j)}(x)$, and the full matrix $\Gamma$ is Metzler.

Define $\tilde{\Gamma}_j := (\mu_{\tau(j), \tau(m)})_{m \in \widetilde{\mathcal{N}}}$, and observe that this is a truncation of the $j$-th row of $\Gamma$.

Then in $\widetilde{\mathcal{I}}$, we have:
\begin{equation}
L_{\tilde{f}_j} \tilde{h}_j + L_{\tilde{g}_j} \tilde{h}_j \tilde{\pi}_j \geq \tilde{\Gamma}_j^\top \tilde{h}.
\end{equation}

Again, the truncated $\tilde{\Gamma}$ remains Metzler as a submatrix of a Metzler matrix, preserving forward invariance of the safe set $C$ via Theorem~2.

\paragraph{Conclusion.}
Since the transferred controllers and certificates in $\widetilde{\mathcal{I}}$ satisfy the same DL-VCLF and DL-VCBF conditions with respect to a substructure-preserving topology and dynamics, Theorems~1 and 2 imply that $\widetilde{\mathcal{I}}$ also satisfies the Co-RWA conditions. Hence, the theorem holds.
\end{proof}

\subsection{Additional Results}
\label{app: additional results}
\begin{figure}[htp]
    \centering
    \begin{subfigure}{0.8\linewidth}
        \centering
        \includegraphics[width=\linewidth]{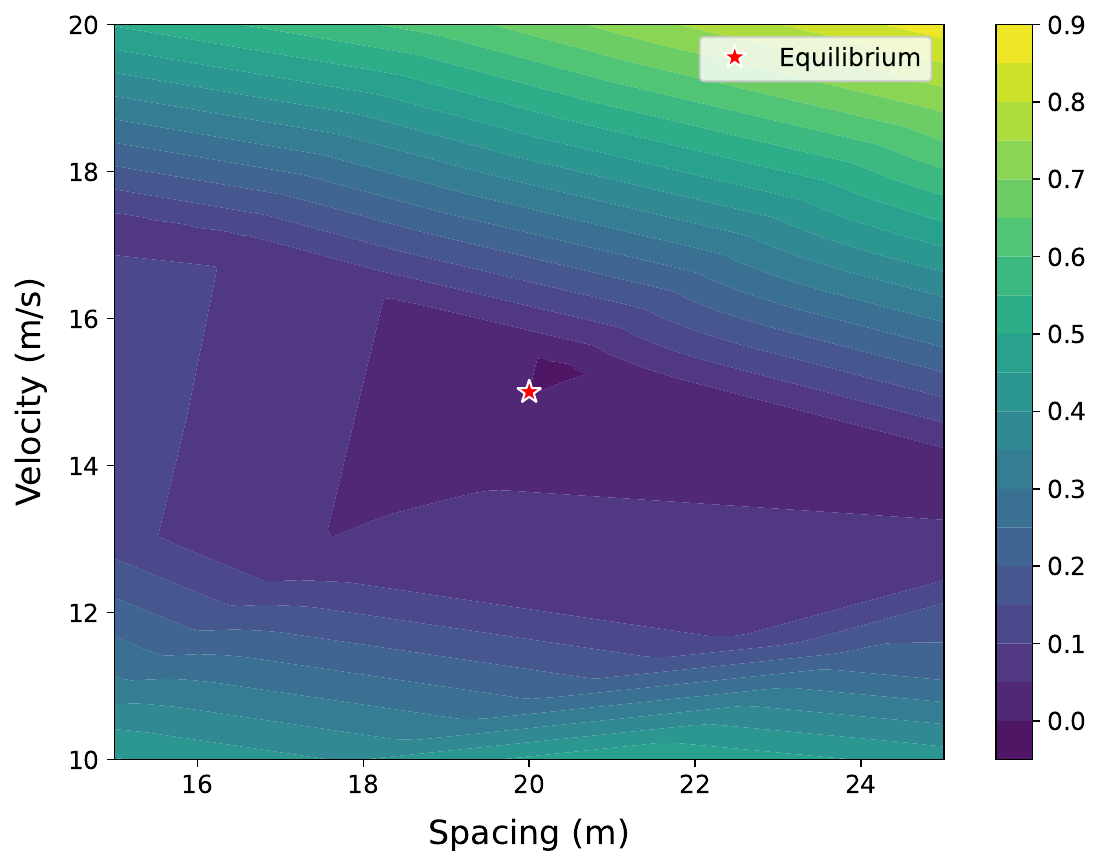}
        \caption{Lyapunov function}
        \label{fig:lyapunov_contour}
    \end{subfigure}
    
    \vspace{1em}

    \begin{subfigure}{0.8\linewidth}
        \centering
        \includegraphics[width=\linewidth]{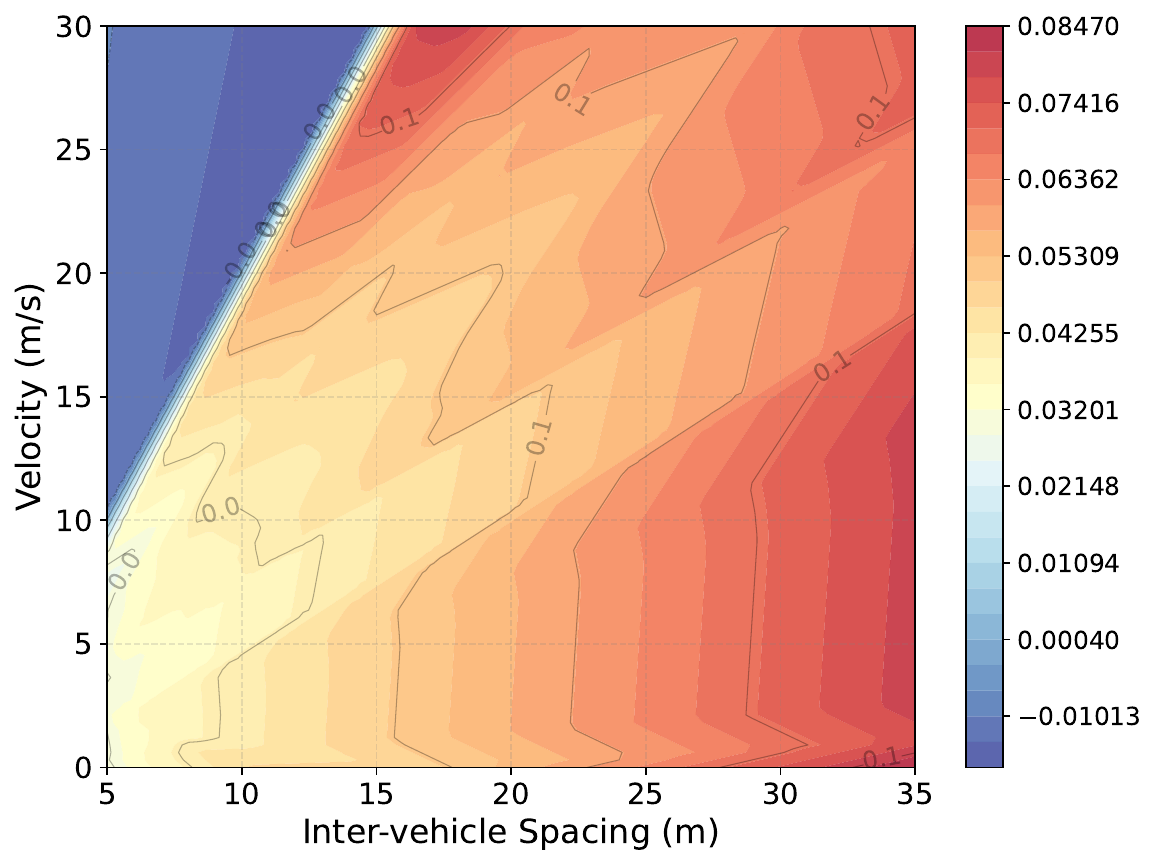}
        \caption{Barrier function}
        \label{fig:barrier_contour}
    \end{subfigure}
    
    \caption{Learned contour plots of Lyapunov and barrier functions for the CAV 1.}
    \label{fig:certificate_contours}
\end{figure}

Figure~\ref{fig:certificate_contours} shows the learned Lyapunov and barrier functions for the CAV system. In Figure~\ref{fig:certificate_contours} (a), the Lyapunov function reaches its minimum at the equilibrium, indicating convergence and stability. The sharp gradient around the equilibrium suggests strong corrective behavior. Figure~\ref{fig:certificate_contours} (b) depicts the barrier function, where the zero-level set separates safe and unsafe states. The function assigns higher values to configurations with larger spacing and moderate velocity, effectively enforcing safety by penalizing close, high-speed states. These plots demonstrate that the learned certificates capture both stability and safety-relevant structures.

Notably, due to the coupling between the Lyapunov and barrier functions in the joint optimization process, the learned functions are not required to be strictly monotonic. Their interaction allows flexibility in the function landscapes, as long as the overall reach–while–avoid specification is satisfied.

\end{document}